\pdfoutput=1
\documentclass{aa}
\usepackage[varg]{txfonts}
\usepackage[]{graphicx}
\usepackage{multirow}
\usepackage{subfig}
\bibpunct{(}{)}{;}{a}{}{,}

\begin{document}
\newcommand{\matisse}{\mbox{MATISSE}}
\title{The AMBRE Project: Parameterisation of FGK-type stars from the ESO:HARPS archived spectra}

\author{M.~De Pascale\inst{\ref{oca}, \ref{eso}, \ref{oapd}, \ref{unipd}}
  \and C.~C.~Worley\inst{\ref{oca}, \ref{ioa}}
  \and P.~de Laverny\inst{\ref{oca}}
  \and A.~Recio-Blanco\inst{\ref{oca}}
  \and V.~Hill\inst{\ref{oca}}
  \and A.~Bijaoui\inst{\ref{oca}}
}
\offprints{Patrick de Laverny, \email laverny@oca.eu }
\institute{Laboratoire Lagrange (UMR7293), Universit\'e de Nice Sophia Antipolis, CNRS, Observatoire de la C\^ote d'Azur, BP 4229,
 F-06304 Nice cedex 4, France\label{oca}
 \and
 Institute of Astronomy, University of Cambridge, Madingley Road, Cambridge CB3 0HA, United Kingdom\label{ioa}
 \and
 European Southern Observatory, Karl-Schwarzschild-Stra\ss e 2, D-85748 Garching bei M\"unchen, Germany\label{eso}
 \and
 INAF - Osservatorio Astronomico di Padova, Vicolo Osservatorio 5, I-35122 Padova, Italy\label{oapd}
 \and
 Dipartimento di Fisica e Astronomia, Vicolo Osservatorio 2, I-35122 Padova, Italy\label{unipd}
}
\date{}
\abstract{
  The AMBRE project is a collaboration between the European Southern Observatory (ESO) and the 
  Observatoire de la C\^ote d'Azur (OCA). It has been established to determine the stellar 
  atmospheric parameters of the archived spectra of four ESO spectrographs.}%
{
  The analysis of the ESO:HARPS archived spectra for the determination of their atmospheric parameters 
  (effective temperature, surface gravity, global metallicities and abundance of $\alpha$-elements over iron) 
  is presented. The sample being analysed (AMBRE:HARPS) covers the period from 2003 to 2010 and is
  comprised of 126\,688 scientific spectra corresponding to $\sim$17\,218 different stars.
}
{
  For the analysis of the AMBRE:HARPS spectral sample, the automated pipeline developed for the analysis of the AMBRE:FEROS 
  archived spectra has been adapted to the characteristics of the HARPS spectra. Within the pipeline, the stellar parameters 
  are determined by the \matisse\, algorithm, which has been developed at OCA for the analysis of large samples of stellar 
  spectra in the framework of galactic archaeology. In the present application, \matisse\, uses the AMBRE grid of synthetic 
  spectra, which covers FGKM-type stars for a range of gravities and metallicities.
}
{
  We first determined the radial velocity and its associated error for the $\sim$15\% of the AMBRE:HARPS spectra, for which 
  this velocity had not been derived by the ESO:HARPS reduction pipeline.
  The stellar atmospheric parameters
  and the associated chemical index [$\alpha$/Fe] with their associated errors 
  have then been estimated for all the spectra of the AMBRE:HARPS archived sample. Based on key quality criteria, we accepted 
  and delivered the parameterisation of 90\,174 (71\% of the total sample) spectra to ESO. These spectra correspond to 
  $\sim$10\,706 stars; each are observed between one and several hundred times. 
  This automatic parameterisation of the AMBRE:HARPS spectra shows that the large majority of these stars are
  cool main-sequence dwarfs with metallicities greater than -0.5~dex (as expected, given that HARPS 
    has been extensively used for planet searches around GK-stars).
}
{}
\maketitle

\section{Introduction}
\label{sec:introduction}

In the last decade, astronomy has entered an era of very large data surveys with the scientific goal
to expand our understanding of the formation and evolution of the Universe. 
In particular, several spectroscopic surveys (for a summary see \cite{RecioBlanco_2012}) are dedicated 
to the study of the Milky Way to comprehend its kinematic and chemical history in detail.
The ESA Gaia mission is the pinnacle of all of these spectroscopic surveys during which its Radial Velocity Spectrometer 
(RVS) will observe tens of millions of stars, for which the radial velocity, atmospheric
parameters, and chemical abundances will be determined.

The analysis of such a large quantity of data using 'by-hand' methods is not feasible on a short time 
scale. This has pushed the astronomical community to develop complex algorithms able to automatically determine the stellar 
parameters for large spectral datasets efficiently and reliably. MATISSE (MATrix Inversion for Spectral SynthEsis) is one 
such algorithm\citep{recio-blanco_automated_2006} that has been developed at the Observatoire de la C\^ote d'Azur (OCA) and 
is part of the automated pipeline that will analyse and parameterise the spectra from Gaia-RVS.

The AMBRE Project \cite[see][]{deLaverny2013}, a collaboration between ESO and OCA, will, in the end have converted the archived spectra provided by 
ESO into a comprehensive spectral library of homogeneously determined stellar
parameters: effective temperature ($T_\mathrm{eff}$), surface gravity ($\log g$), metallicity ([M/H]), and the abundance of 
$\alpha$-elements versus iron ([$\alpha$/Fe). These quantities will be made publicly available to the international scientific 
community, as advanced data products via the ESO archive. The AMBRE Project has two other main objectives: first, to rigorously test
 MATISSE on large spectral datasets over a range of wavelengths and resolutions that include those of the Gaia-RVS and, second, 
to produce a chemo-kinematical map of the Galaxy using the combined ESO archive samples to unravel galactic formation and evolution.

The first part of the AMBRE project consisted of the analysis of the FEROS archived spectra. This has been presented in
\citet{worley_ambre_2012}, and the parameters are now publicly available. The present paper reports the work carried out on 
the HARPS archived spectra provided by ESO. In Sec.~\ref{sec:AMBRE-proj}, we describe the dataset and its properties, 
with the adaptation of the AMBRE:FEROS analysis pipeline to the AMBRE:HARPS sample and the derivation of the 
radial velocities. Section~\ref{sec:internal-err} we exaplain how we used a sample of 
stars with repeated observations to determine the internal error.
Section~\ref{sec:external-err} describes the external errors estimates by the 
comparison of key samples with literature parameter values. Finally, Sec.~\ref{sec:results} presents the parameterisation 
of the accepted AMBRE:HARPS spectra with their delivery to the ESO archive, and we conclude in Sec.~\ref{sec:summary} with a short 
summary.

\section{The AMBRE analysis of the HARPS spectra}
\label{sec:AMBRE-proj}

The High Accuracy Radial velocity Planet Searcher \citep[HARPS][]{mayor_setting_2003} is 
a fiber-fed, cross-dispersed echelle spectrograph that was built by a consortium of four institutes:
the {\it Observatoire de Gen\`eve}, the {\it Observatoire de Haute Provence}, the {\it Universit\"at Bern}, and 
the {Service d'A\'eronomie} of CNRS in collaboration with ESO. It was installed and commissioned on the
ESO 3.6m Telescope at La Silla, Chile, in 2003\footnote{\url{http://www.eso.org/sci/facilities/lasilla/instruments/harps/index.html}}. 
The HARPS high resolution ($\mathrm{R} \simeq 120\,000$) spectra and the long term instrument stability ensure a 
radial velocity accuracy of about 1~m~s$^{-1}$
\citep{harpsman2011}, making HARPS the prime facility for exoplanet hunting.

The AMBRE analysis of the HARPS spectra comprises of the spectra observed from October 2003 to October 2010. They have been 
homogeneously reduced by ESO with the HARPS pipeline and made publicly available through the ESO archives.
This sample was delivered to OCA from the ESO archive department and it includes calibration and science spectra. 
The ESO:HARPS pipeline produces different types of science spectra: the extracted 2-Dimensional spectra, where each line 
contains the extracted flux from one spectral order; the extracted 1-Dimensional spectra, which contains the re-binned 
and merged spectral orders; and the radial velocity cross correlation function (CCF), which is computed between each order and a 
template mask. For the AMBRE analysis, we used the 126\,688 extracted 1-Dimensional science spectra. About 85\% of these 
spectra have a corresponding CCF spectrum and, thus a radial velocity ($V_\mathrm{rad}$) estimate. For the 
remaining 15\%, we calculated the $V_\mathrm{rad}$ using the AMBRE automatic program \citep[see][Sec.~4.2 and 
Sec.~\ref{sec:rad-vel} of the present paper]{worley_ambre_2012}.

\begin{figure}
  \resizebox{\hsize}{!}{\includegraphics{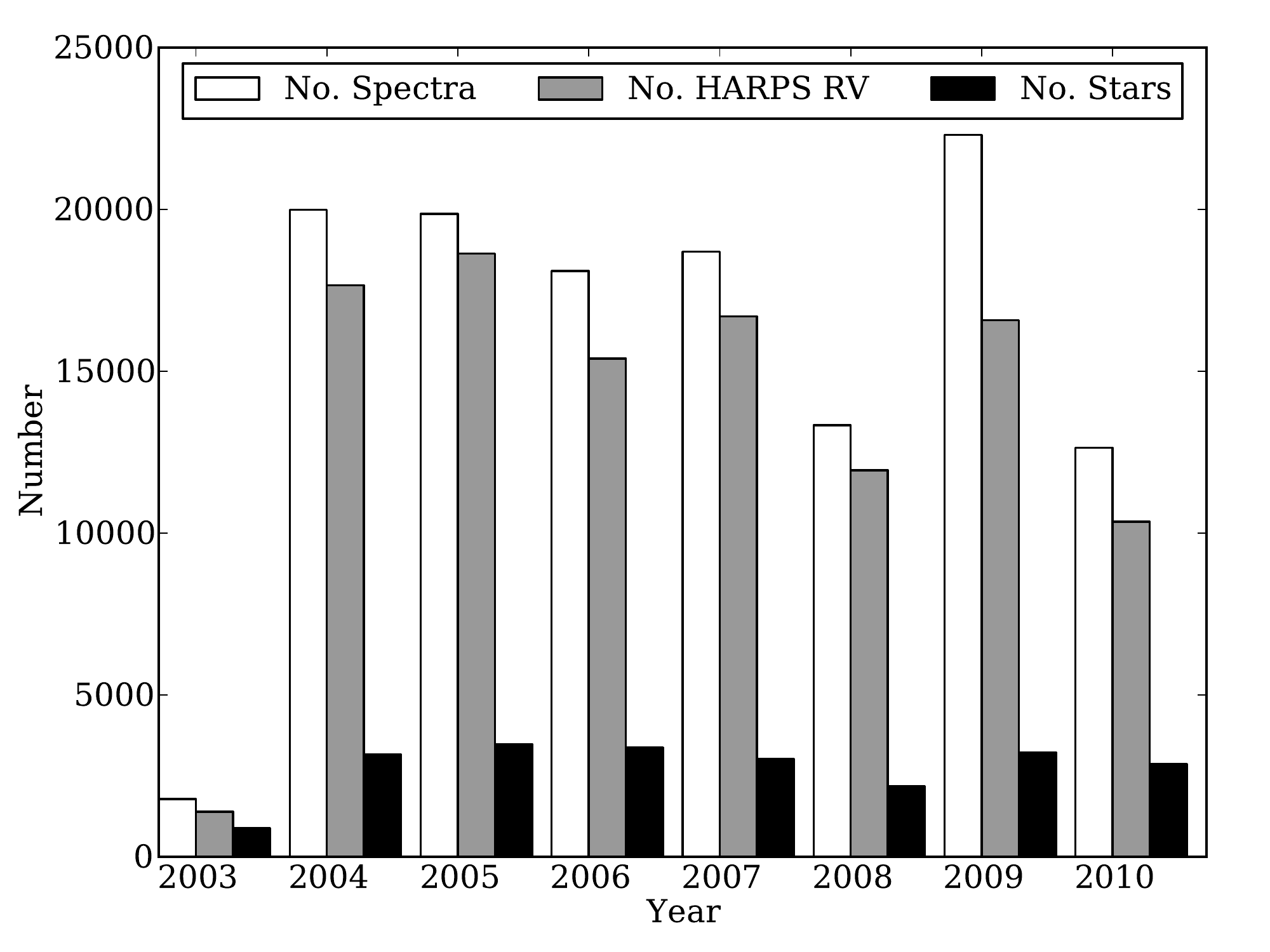}}
  \caption{Number of HARPS spectra per year analysed by the AMBRE Project (in white). 
    The number of spectra for which a $V_\mathrm{rad}$
    was calculated by the ESO:HARPS pipeline are shown in gray.
    The number of different stars observed per year are shown in black.}
  \label{fig:harps-obs}
\end{figure}

This AMBRE:HARPS sample of 126\,688 spectra is composed of several thousands of distinct stars with some of them being 
observed a few tens of times. We have found that the object ID available in the file header was not always a
reliable indicator; thus, we have determined the total number of distinct stars included in this sample by performing 
a coordinate matching analysis whereby a maximum distance radius was imposed.
Within a radius of $r \simeq 5 \arcsec$, we obtain 17\,218 
distinct stars. The counts of the number of science spectra and, thus, distinct objects that
have been observed by HARPS in the given period are reported in Fig.~\ref{fig:harps-obs} with the number of 
spectra with $V_\mathrm{rad}$ calculated by the ESO:HARPS pipeline. 

\begin{figure}
  \centering
  \resizebox{\hsize}{!}{\includegraphics{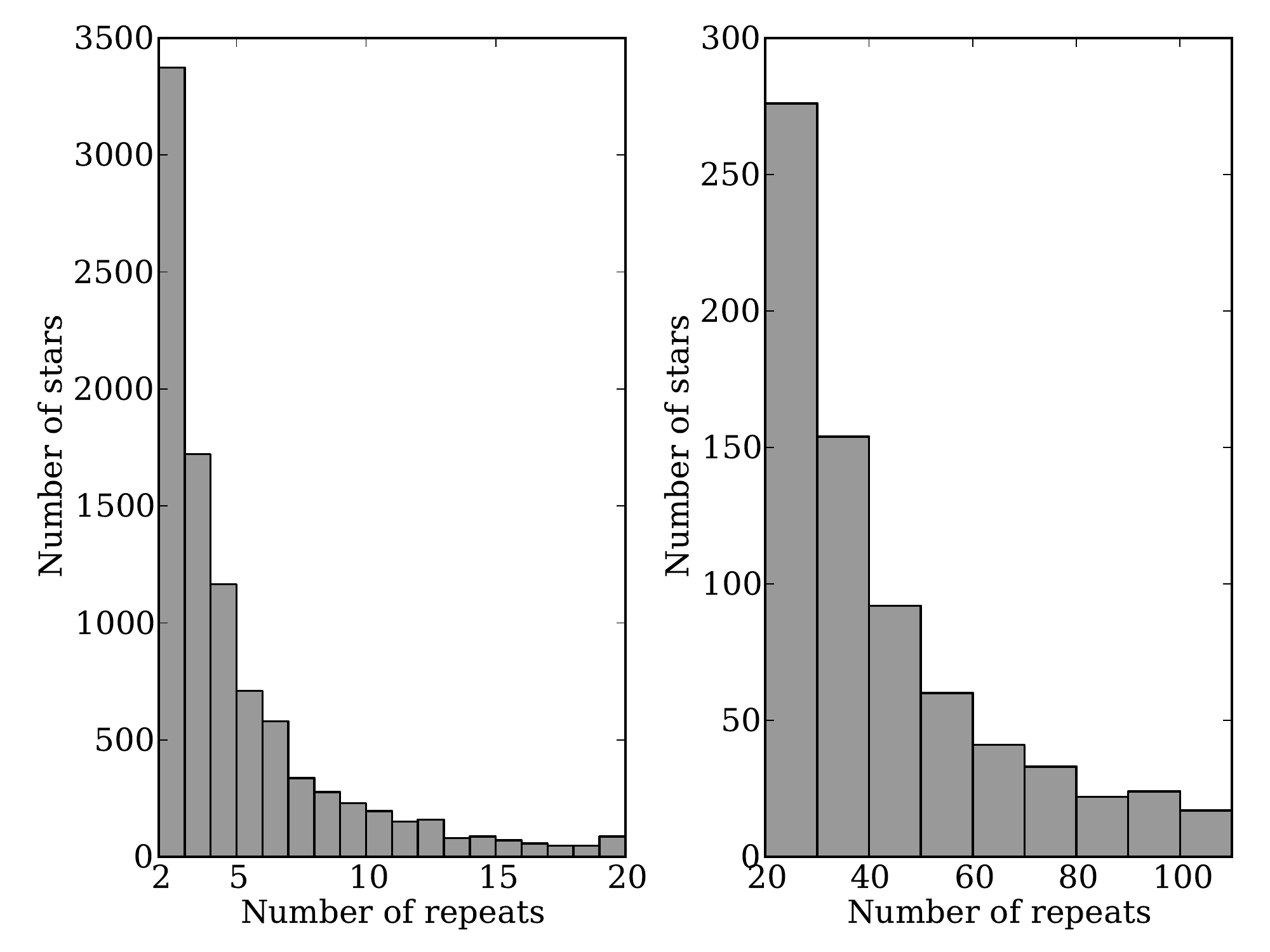}}
  \caption{Approximate number of stars in the sample as a function of the number of observations. The total 
  number of observed stars is 17\,218. 
\emph{Left panel}: Distribution of the repeats for stars observed between 2 and 20 times.
\emph{Right panel}: Similarly for stars observed more than 20 times (see text for more details).}
  \label{fig:spectra-nobs}
\end{figure}

Figure~\ref{fig:spectra-nobs} shows the distribution of the repeated observations, which shows the number of different spectra 
available for the same star within the AMBRE:HARPS sample. About $\sim 40$\% of the stars were observed only once, and 
$\sim 95$\% of the sample less than 20 times. 
Within the search radius of $\sim$5~$\arcsec$, we have found that only 182 targets have been observed more then 110 times
 with a maximum of 1\,211 repeated observations for Procyon AB binary system. 

\subsection{Radial velocity}
\label{sec:rad-vel}

To analyse the observed spectra with \matisse, it is necessary to correct them by the 
radial velocity of the star. Since the main scientific goal of HARPS is the search for exoplanets 
by measuring variations in the radial velocity of the host star, the data reduction pipeline (DRS) 
of HARPS determines the radial velocities with extremely high accuracy ($\sim$1~m~s$^{-1}$). Radial 
velocity error estimates are also provided. Given the large number of spectra to be analysed
 and for the purpose of homogeneity with the ESO archives, we chose to adopt the 
radial velocity and the associated error provided in the header of the delivered reduced spectra
 when available. 

However, as previously mentioned, among the sample of spectra delivered to OCA, approximately $15\%$
  do not have a HARPS:DRS radial velocity. Almost all of this subsample of 
spectra have a radial velocity set to a default value from the DRS, while a tiny fraction 
($\sim 2\%$) have calculated values that are larger than 500~km~s$^{-1}$ in modulus. This 
indicates that the DRS radial velocity routine had most likely not converged. We therefore 
calculated the $V_\mathrm{rad}$ for this subsample of spectra using the AMBRE pipeline 
\citep[see][Sec.~4.2]{worley_ambre_2012}. 
Briefly, the radial velocity routine performs a cross-correlation between each spectrum and 
a set of 56 synthetic masks specifically computed for the AMBRE:HARPS sample. 
For each spectrum, the output is therefore a set of 56 radial velocity 
determinations with associated errors; the radial velocity with the lowest error has been adopted.
We point out that a final check on the validity of the adopted $V_\mathrm{rad}$ has been performed 
once the analysed spectra have been parametrized. We indeed always found that the 
atmospheric parameters estimated for a HARPS spectrum and those of the adopted mask
agree with each other.

The AMBRE $V_\mathrm{rad}$ procedure also gives an estimate of the uncertainty associated with the
 derived radial velocity. To check the consistency of these results with the HARPS $V_\mathrm{rad}$, 
we also determined the radial velocity for a sample of spectra having a $V_\mathrm{rad}$ computed by
 HARPS:DRS. For that purpose, we considered a sample of $\sim$17\,000 HARPS spectra, which were
 observed in 2004. 

\begin{figure}
  \centering
  \resizebox{\hsize}{!}{\includegraphics{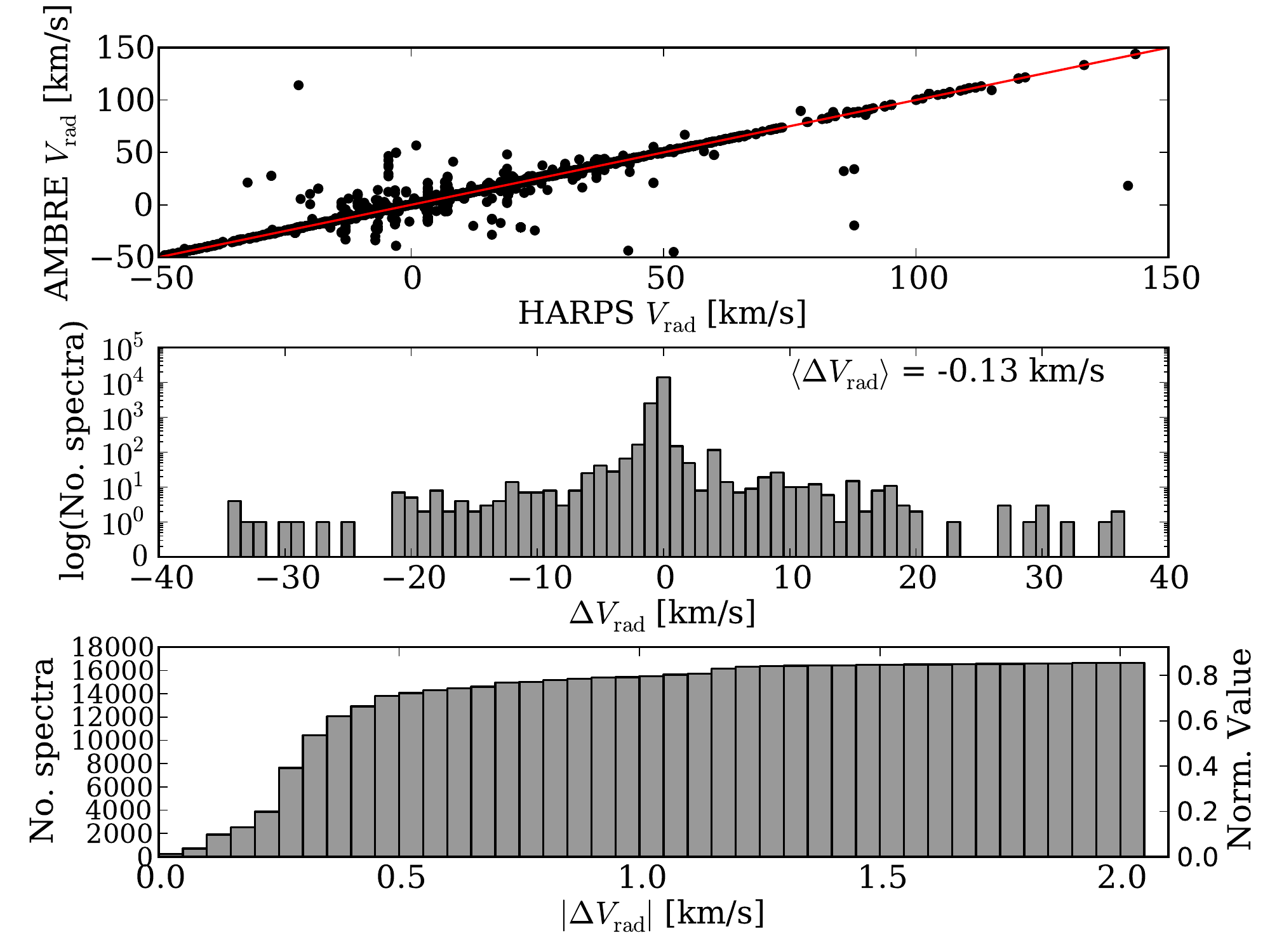}}
  \caption{\emph{Top panel}: Comparison between the radial velocity as calculated from the AMBRE 
    radial velocity program and from the HARPS pipeline for the 2004 sample ($\sim$ 17\,000 spectra). 
     The distribution is Gaussian.
    \emph{Middle panel}: Distribution of $\Delta V_\mathrm{rad}$ between HARPS and AMBRE radial 
    velocities using a logarithmic scale. \emph{Bottom panel}:     Cumulative distribution of the 
    $|\Delta V_\mathrm{rad}|$ in bins of 0.05~km~s$^{-1}$. Almost 70\% of the spectra have a 
    $|\Delta V_\mathrm{rad}|$ smaller than 0.6~km~s$^{-1}$. Moreover, almost 90\% of the spectra are 
    found between $\Delta V_\mathrm{rad} = \pm$1~km~s$^{-1}$.}
  \label{fig:rv-comparison}
\end{figure}

This comparison between the HARPS:DRS $V_\mathrm{rad}$ and the $V_\mathrm{rad}$ derived by AMBRE is 
shown in Fig.~\ref{fig:rv-comparison}. The distribution of the residuals between the two estimates 
in the bottom panel of Fig.~\ref{fig:rv-comparison}, shows a Gaussian distribution
centered close to zero ($\langle V_\mathrm{rad}^\mathrm{HARPS} - V_\mathrm{rad}^\mathrm{AMBRE} 
\rangle = -0.13$~km~s$^{-1}$). Additionally, 89\% of the spectra have an absolute value of the 
residual, which is smaller than 1~km~s$^{-1}$, whereas 96\% of them have a difference smaller than 
2~km~s$^{-1}$. This test confirms the high enough consistency between the HARPS $V_\mathrm{rad}$ and 
the AMBRE $V_\mathrm{rad}$, such that the use of the AMBRE $V_\mathrm{rad}$ for those spectra 
without a HARPS $V_\mathrm{rad}$ stills provide a homogeneous parameterisation analysis across 
the entire sample.
Moreover, with reference to \citep[Fig.~11]{worley_ambre_2012}, we note that an error of 
2~km~s$^{-1}$ on the radial velocity has very little effect on the determination of the atmospheric 
parameters. 

In summary, we therefore adopted the radial velocity and associated error 
provided by HARPS:DRS for the AMBRE analysis whenever possible. When these quantities were not 
available, we calculated them using the AMBRE procedure that has been shown to be consistent with 
the HARPS:DRS results.

\subsection{The AMBRE:HARPS parameterisation pipeline}
\label{sec:pipeline}
For the analysis of the HARPS spectra, we started from the pipeline that was developed for the 
AMBRE:FEROS analysis, which is described in detail in \citet{worley_ambre_2012}. Due to the 
inherent differences between the two instruments in configuration (principally resolution and 
spectral range) and in the ESO reduction pipeline products (the radial velocity is provided for 
each HARPS spectra, for instance), the AMBRE:FEROS pipeline was adapted to obtain an 
optimal analysis for the HARPS spectra. These modifications are highlighted in the following 
subsections.

\subsubsection{Adaptation of the HARPS spectra for the AMBRE analysis}
\label{sec:adapt-feros}
The parameterisation of the stellar spectra in AMBRE is performed by  a
kind of comparison of the observed 
spectra with a library of synthetic spectra using the \matisse\, algorithm 
\citep{recio-blanco_automated_2006}. We remind that \matisse\, is a local 
multi-linear regression method. It acts as a
projection method in the sense that the input observed spectra are 
projected onto a set of vectors derived during a learning phase of \matisse .
These vectors are a linear combination of reference spectra
(i.e. the synthetic spectra described below)
and could be roughly viewed as the derivatives of these 
spectra with respect to the different stellar parameters.
We point out that we adopted exactly the same version of \matisse\,
for the analysis of the FEROS and HARPS spectra (see comment in Sect.~\ref{sec:final-param}).

Moreover, the same grid of synthetic spectra, the AMBRE grid, has also been 
adopted. Shortly, this grid of 
$\sim$17\,000 high-resolution synthetic spectra has been computed from MARCS model atmospheres
\citep{gustafsson_2008}, taking into account the most complete atomic and molecular 
linelists. The spectra cover the whole optical domain for cool to very cool stars of any 
luminosity (from dwarfs to supergiants) with metallicities varying from 10$^{-5}$ to 10 times 
the Solar value. Large variations in the chemical composition of the $\alpha$-elements
with respect to iron have also been considered. As in the MARCS models,
a constant microturbulent velocity has been adopted for dwarfs (1 km/s)
and giants (2 km/s). More details on the properties of the 
AMBRE grid and how it has been computed can be found in \citet{de_laverny_ambre_2012}.

The wavelength coverage of the AMBRE synthetic spectra grid goes from 300~nm to 1\,200~nm, which is
the whole optical wavelength domain. We, thus, were able to select only those wavelengths corresponding 
to the HARPS wavelength domain that were useful for the analysis. HARPS disperses light on 68 
orders covering the spectral range between 378~nm and 691~nm with a gap from 530~nm to 533~nm due 
to the two CCDs that form the detector system of the instrument. Since we analysed the extracted 
1-Dimensional spectra, we first discarded the blue and red edges of the CCDs where the 
signal-to-noise (S/N) can be significantly lower with respect to the remainder of the spectrum.

Then, from the two wavelength domains defined by the two CCDs, we rejected sections that contained 
sky absorption and telluric features. In addition, we rejected the very broad \ion{Ca}{II} H \& K 
lines, since they can be poorly synthesised for some parameter combinations and they are difficult 
to normalise automatically.
The accepted wavelength regions for the AMBRE:HARPS analysis are listed in 
Table~\ref{tab:harps-wave-region}.

\begin{table}
  \caption{Selected HARPS wavelength domains for the AMBRE analysis.}
  \label{tab:harps-wave-region}
  \centering
  \begin{tabular}{cll}
    \hline\hline
    \multirow{2}{*}{Region} & \multicolumn{1}{c}{$\lambda$ Min} & \multicolumn{1}{c}{$\lambda$ Max} \\
    & \multicolumn{1}{c}{(nm)} & \multicolumn{1}{c}{(nm)} \\
    \hline
    1   & 400.0 & 500.0 \\
    2   & 513.0 & 530.0 \\
    3   & 552.0 & 566.0 \\
    4   & 575.0 & 580.0 \\
    5   & 600.0 & 627.0 \\
    6   & 635.0 & 645.0 \\
    7   & 661.0 & 685.0 \\
    \hline
  \end{tabular}
  \tablefoot{These listed wavelength intervals are not (or are weakly) polluted by absorption and 
    telluric features and do not contain the gap present between the two CCDs, the lowest S/N 
    regions, nor the regions of the \ion{Ca}{II} H and K lines.}
\end{table}

A further refined selection of the wavelength ranges was then performed by comparing the observed 
normalised spectra of the Sun and Arcturus line by line 
(the two stars are representatives of standard dwarf and giant stars and taken from \citep{SolarAtlasHinkle,ArcturusAtlasHinkle}) 
with the corresponding synthetic spectra in the AMBRE grid. Lines at matching rest wavelengths were rejected when 
the percentage difference between their two fluxes was larger then a fixed threshold. After testing, we selected a 
threshold of 0.05\% for the Sun and 0.15\% for Arcturus. Both thresholds had to be met for the spectral line to be 
included. These thresholds removed obvious mismatches between the observed and synthetic spectra. A higher threshold 
was set for Arcturus since the underlying physics of giants is less understood than it is for the Sun and hence 
the spectra of giants are not as well synthesised. This threshold is also allowed for greater inaccuracies in normalisation between 
the Arcturus and its corresponding synthetic spectrum.

This procedure allowed us to define the final list of selected wavelengths for the AMBRE:HARPS parameterisation. 
It consists of $\sim 500$ intervals sampling the previously selected HARPS domains reported in 
Table~\ref{tab:harps-wave-region} and spanning a total range of about 147~nm. Using this final list of wavelength intervals, 
we extracted the corresponding ranges from the grid of synthetic spectra resulting in the AMBRE:HARPS synthetic spectra grid.

As the resolution and pixel sampling of HARPS are very high, the computing time required to analyse the original 
spectra is also correspondingly high. As for the AMBRE:FEROS analysis, the resolution and pixel sampling can be 
lowered to optimise computing time but without sacrifying the key spectral informations with 
the goal being to keep an as good as possible accuracy on the derived stellar parameters. This has been explored 
in our previous AMBRE:FEROS analysis and we have found that a resolution $R \sim 15\,000$ was sufficient
to achieve the required accuracy.

The HARPS spectra have a constant resolution ($R = \lambda/\Delta \lambda$) and hence a varying $\Delta \lambda$ with 
$\lambda$ (where $\Delta \lambda$ is the full-width-at-half-maximum (FWHM) of the spectral feature at $\lambda$). This 
is contrary to the synthetic spectra, which have been computed without any instrumental nor macroturbolence profiles,
and a constant wavelength sampling of 0.001~nm.

By degrading (or convolving) both the synthetic and observed spectra to a lower resolution and mapping the observed 
FWHM profile onto the synthetic FWHM profile, computation time can be decreased and the comparison between the two 
sets of spectra is then consistent. The convolution was performed by ``smoothing'' the spectra with a Gaussian for  
which the FWHM was greater (and therefore of lower resolution) than that of HARPS. 

The synthetic grid was convolved with a Gaussian of FWHM=0.02218~nm to produce a synthetic grid with resolution less 
than that of HARPS. 

For each observed spectrum, the measured variation of $\Delta \lambda$ as a function of $\lambda$ was interpolated 
to a linear function providing a uniformly increasing FWHM profile for each spectrum (where possible).
Similarly, for a subsample of synthetic spectrum in the grid, their FWHM profile was also measured by the same 
procedure, confirming the constant FWHM profile. The mean value from the combination of all of these synthetic 
FWHM profiles was found to be FWHM$_\mathrm{mean} = 0.022$~nm, as expected from the grid convolution, confirming 
that our procedure is valid. This was taken to be the nominal synthetic grid FWHM.

To map the convolution of the observed onto the synthetic for each bin in $\lambda$, the FWHM of the smoothing 
Gaussian was calculated using FWHM$_\mathrm{mean}$ and the observed was linearly interpolated FWHM for that bin. This 
resulted in the convolved observed spectra having a constant FWHM profile of the same resolution as the convolved 
synthetic grid.

The final wavelength sampling was chosen to fulfill the Shannon criterion resulting in a sampling of 
0.0085~nm/pixel. The same sampling was also applied to the synthetic spectra of the AMBRE:HARPS 
grid from which we generated the learning functions of \matisse\, to parameterise the observed spectra.
We remind that the broadening caused
by the micro-turbulent velocity is dominated by the adopted one for the analysis.
In consequence, this parameter is not estimated by the pipeline that relies
on the adopted constant micro-turbulent velocity of the synthetic grid.
We also do not derive the projected rotational velocity and reject
every spectra having a too large line-broadening that could affect
our parameter estimates (see our discussion in \cite{worley_ambre_2012}).

\subsubsection{Spectral processing A, B and C}
\label{sec:spa}
In \citet[][Sec.~4 and Fig.~4]{worley_ambre_2012}, a detailed description of the several
steps that are part of the AMBRE pipeline with a graphical representation of the process
 is reported. We review the three main stages of the AMBRE pipeline here. 
We refer the reader to the AMBRE:FEROS analysis for more detail.

First, we remind that the HARPS spectra delivered to us by ESO are the products of the 
standard data reduction pipeline. As a consequence, some degradations
could still be present in some spectra as, for instance, some possible contaminations 
by the wavelength calibration lamp for the low S/N ones (about
7\% of the whole AMBRE:HARPS sample has S/N $<$ 20).
To keep the analysis of the whole sample as homogeneous as possible,
 we preferred to ignore such effects in the following.

In spectral processing A (SPA), the observed spectra are prepared for radial velocity computation 
and spectral FWHM measurements, and a first quality check is performed identifying noisy or problematic 
spectra. Each spectrum is sliced in the wavelength regions defined above and roughly normalised to unity. 
Previously in \citet{worley_ambre_2012}, the spectral FWHM was measured during the next stage. However, 
as this was heavy in computing time, it has been developed as a standalone routine for AMBRE:HARPS 
and AMBRE:UVES that can be run in parallel to the radial velocity routine on the normalised spectra of SPA. 
The radial velocity program was used only on the spectra for which the HARPS radial velocity was 
not available, and the radial velocity testing sample (see Section~\ref{sec:rad-vel}). During SPA, we calculated
the S/N of the spectra (since it was not reported in the header of the reduced spectra), as done in the AMBRE:FEROS analysis.

Spectral processing B (SPB) is the stage at which the first estimate of the stellar parameters is made using 
\matisse. The observed spectra are, thus, normalised and convolved to be consistent with the 
AMBRE:HARPS synthetic grid. At this stage, there is still no determination of the atmospheric parameters;
thus, the spectra are normalized to unity. 
A key point is also the flagging and rejection of problematic spectra; spectra 
flagged as ``bad normalized'', ``noisy'' or ``missing wavelength'' are rejected before the subsequent stage.
We point out that, contrary to the SPB of \citet[][]{worley_ambre_2012}, 
no iteration were performed for the HARPS spectra in SPB since the SPB estimate of the stellar
parameters was always close to the final solution.

Finally, rather than normalisation to unity, the normalisation of each spectrum in Spectral Processing C (SPC), is
performed on a synthetic spectrum to better represent the continuum placement of the star. In the first instance, 
the normalisation is performed on the synthetic spectrum generated at the stellar parameters of the solution found in SPB. 
The resulting normalised spectrum is then analysed in \matisse, which provides yet another solution and 
corresponding synthetic spectrum. Thus, SPC consists in iterating between normalisation and parameterisation, 
ultimately converging on the final stellar parameters, the final normalised spectrum and the final synthetic 
spectrum. A quality flag is produced, based on a $\chi^2$ fit between the observed spectrum and synthetic spectrum 
at the determined stellar parameters.

\subsubsection{Rejection criteria}
\label{sec:rejection}
\begin{figure}[h]
    \centering
    \resizebox{\hsize}{!}{\includegraphics{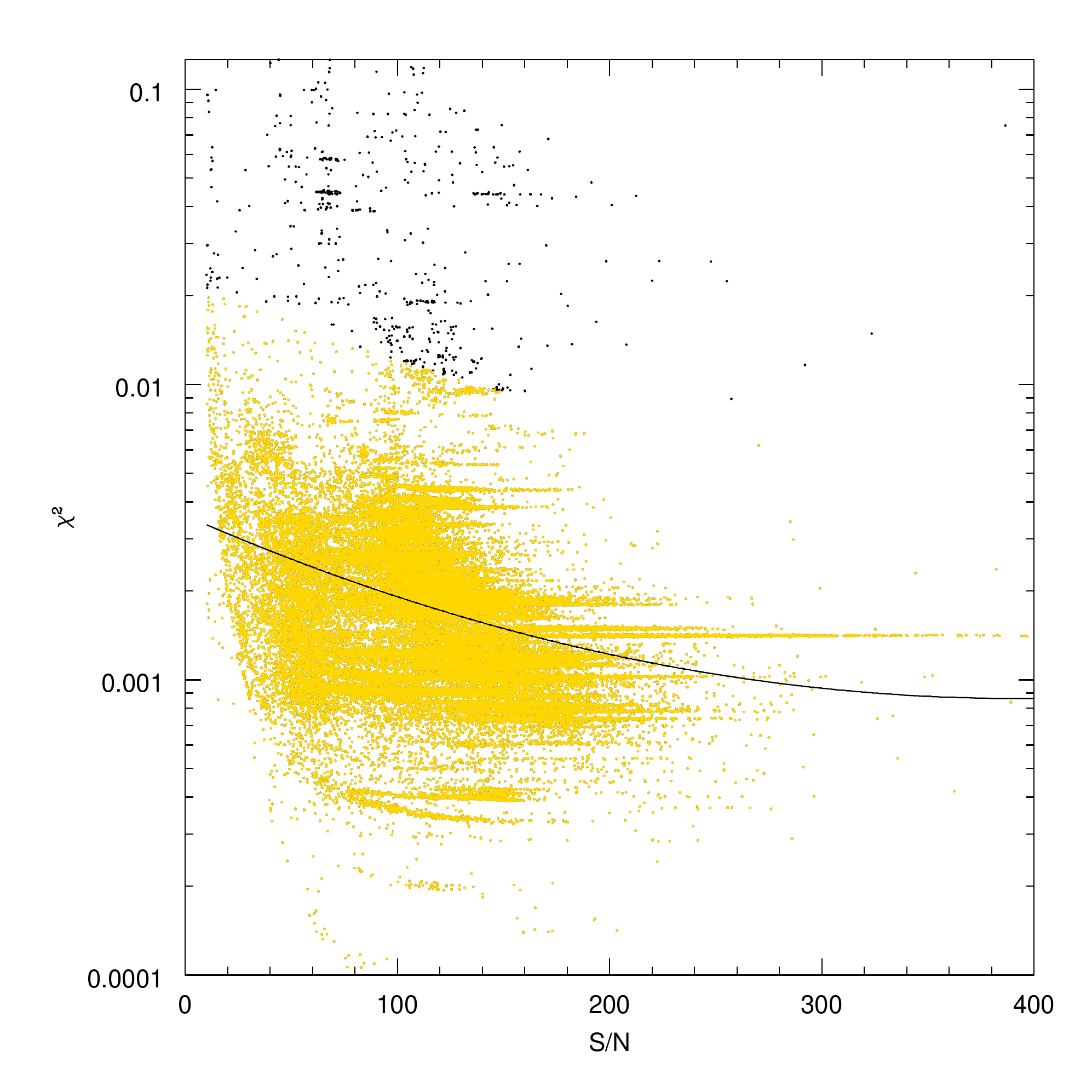}}
    \caption{Spectra selection on the basis of the $\chi^2$ quality criterion as function of S/N for stars with 
      5\,000~K $< \mathrm{T}_\mathrm{eff} \leq$~6\,500~K. The solid line traces the second degree polynomial fit to those 
      spectra. The yellow points are the selected spectra, starting from a threshold of three times the standard 
      deviation above the fit.} 
    \label{fig:chi2_vs_SNR}
  \end{figure}

\newcommand{\acceptedspectra}{93\,116}
\newcommand{\rvrej}{14\,105}
\newcommand{\rvrejper}{11\%}
\newcommand{\rvfwhmrej}{5\,199}
\newcommand{\rvfwhmrejper}{4\%}
\newcommand{\chisquaresnrrej}{6\,340}
\newcommand{\chisquaresnrrejper}{5\%}
The final parameters delivered to the ESO archive are a subsample of the entire set of parameters estimated by our 
pipeline due to quality selection based on several rejection criteria.
As in \citet[][Sec.~7]{worley_ambre_2012}, the first set of rejection criteria that were applied were
based on the radial velocity error ($\sigma_{V_\mathrm{rad}}$), the FWHM of the $V_\mathrm{rad}$ CCF, the S/N, and 
the quality of the fit of the normalised spectra to the synthetic spectra ($\chi^2$):

\begin{itemize}
\item Following the rejection procedure of the AMBRE:FEROS data, all HARPS spectra with $\sigma_{V_\mathrm{rad}} > 10$~km~s$^{-1}$
  were rejected as corresponding to large uncertainties in the parameter determination, see 
  \citet[][Sec.~3 \& 5.2 and Fig.~11]{worley_ambre_2012}. Specifically, for such large errors on the radial
velocity, the uncertainties would be greater than 
$\sim 120$~K in $T_\mathrm{eff}$, 
  $\sim 0.4$~dex in $\log g$, $\sim 0.35$~dex in [M/H] and $\sim 0.17$~dex in [$\alpha$/Fe].
  This step resulted in the rejection of \rvrej\, spectra (\rvrejper\, of the total sample). 
\item With reference to \citet[][Sec.~7.2.3]{worley_ambre_2012}, all spectra with a FWHM of the CCF larger than 20~km~s$^{-1}$ 
  (hot/fast rotating stars) were rejected. Such a threshold value was also chosen on the basis of the results 
    obtained by \citet{gazzano2010}.
  It excluded another \rvfwhmrej\, spectra from the final results (\rvfwhmrejper\, of the total).
\item We also excluded every spectra having a S/N smaller than 10, since the parameterisation 
of such spectra are associated with rather large internal errors (see Sec.~\ref{sec:internal-err}).
\item Finally, to apply a rejection criterion based on the S/N of the HARPS spectra 
with the quality of their parameterisation, we first divided the remaining set 
  of spectra into three different temperature domains: hot stars (T$_\mathrm{eff} > 6\,500$~K), warm stars 
  (5\,000~K $< \mathrm{T}_\mathrm{eff} \leq$~6\,500~K) and cool stars (4\,000~K $\leq \mathrm{T}_\mathrm{eff} 
  \leq$~5\,000~K).
  For each of the three temperature domains, the S/N threshold used to 
  reject spectra was determined by fitting a second degree polynomial to the distribution of $\chi^2$ 
  as a function of the S/N. This $\chi^2$ corresponds to the sum of the squared differences between the synthetic
  and observed fluxes performed at every pixel:
  \begin{equation}
    \label{eq:chi2}
    \chi^2 \sim \sum_\mathrm{px}\left(F_\mathrm{syn} - F_\mathrm{obs}\right)^2.
  \end{equation}
  We investigated the effect of applying different thresholds in the three temperature domains by looking at 
  the distribution of the rejected spectra in the HR diagram. The optimal selection for cool stars was obtained by 
  retaining all the spectra below 0.5 times the standard deviation above the fit for cool stars. 
  Similarly, we kept the warm star spectra by having a $\chi^2$ smaller than three times the standard 
  deviation above the fit (see Fig.~\ref{fig:chi2_vs_SNR} and Table~\ref{tab:chi2-flag}) and 
  finally, every hot star spectra located below the fit in the S/N-$\chi^2$ space. This last 
  stage resulted in the rejection of \chisquaresnrrej\, spectra (\chisquaresnrrejper\, of the total).
\end{itemize}

As last rejection criterion, we applied the restrictions imposed by the synthetic grid boundaries, 
particularly excluding stars cooler then 4\,000~K, since the determination of gravity 
was uncertain in the temperature range between 3\,000~K and 4\,000~K. The adopted grid limits are as follows:

\begin{itemize}
\item 4\,000 K $\leq$ T$_\mathrm{eff}$ $\leq$ 7\,625 K;
\item 1 dex $\leq$ $\log g$ $\leq$ 5 dex;
\item -3.5 dex $\leq$ [M/H]          $\leq$ 1   dex;
\item -0.4 dex $\leq$ [$\alpha$/Fe]  $\leq$ 0.4 dex \quad if [M/H] $\geq$ 0.0 dex;
\item -0.4 dex $\leq$ [$\alpha$/Fe]  $\leq$ 0.8 dex \quad if -1.0 dex $<$ [M/H] $<$ 0.0 dex;
\item  0.0 dex $\leq$ [$\alpha$/Fe]  $\leq$ 0.8 dex \quad if [M/H] $\leq$ -1.0 dex.
\end{itemize}

After the application of these rejection criteria, stellar parameters for \acceptedspectra\, spectra were accepted
(i.e. $\simeq$73\% of the initial sample). The stellar parameters
derived for this AMBRE:HARPS sample of spectra are described in Sec.~\ref{sec:results}.

\section{Internal error analysis}
\label{sec:internal-err}
\begin{figure}[h]
  \centering
  \resizebox{\hsize}{!}{\includegraphics{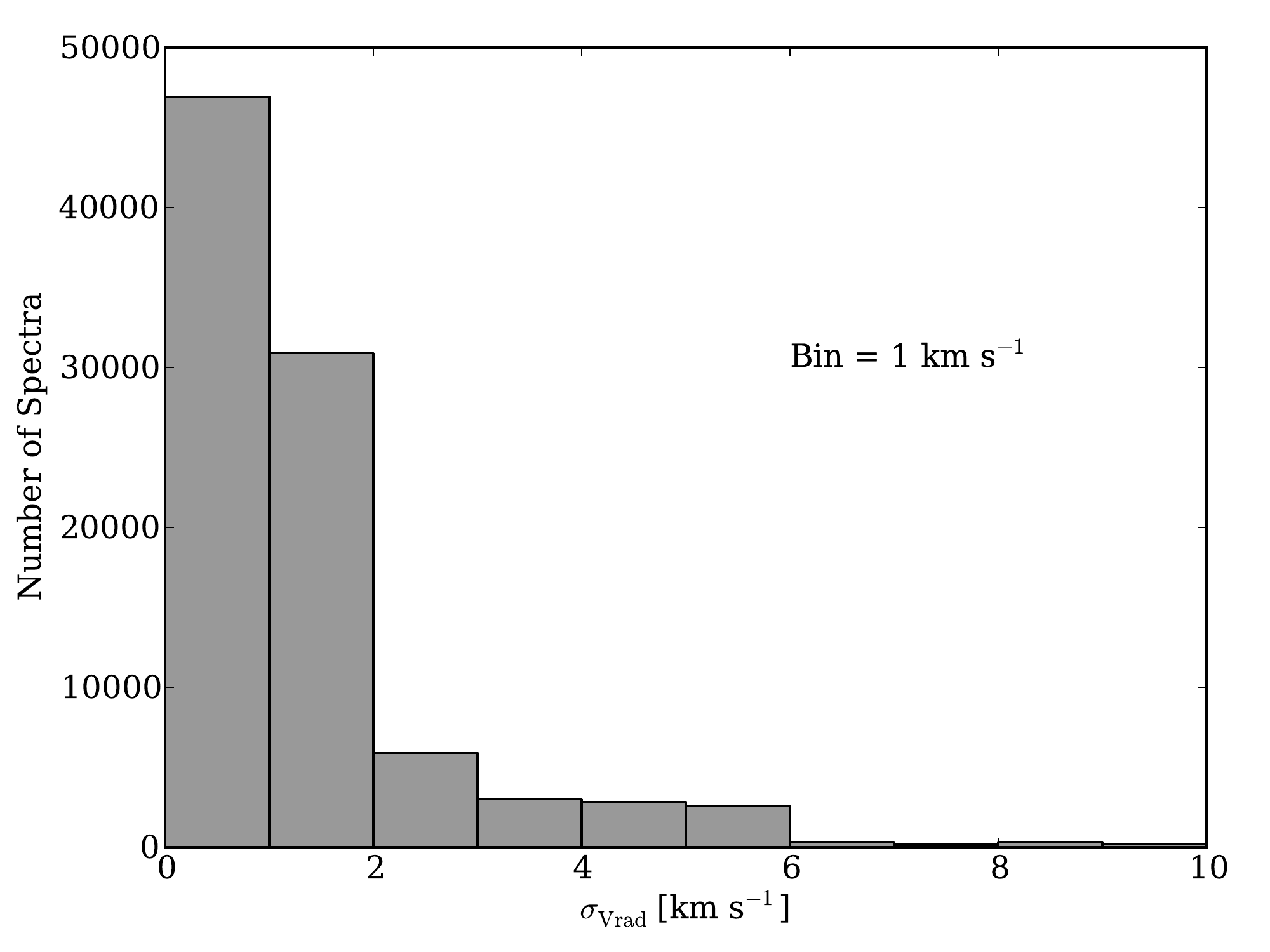}}
  \caption{Histogram of measured $V_\mathrm{rad}$ uncertainty for each of the HARPS spectra that satisfied the
    rejection criteria. The errors come from the HARPS:DRS pipeline or the AMBRE radial velocity determination 
    routine, as appropriate.}
  \label{fig:vrad-err-hist}
\end{figure}

\newcommand{\repeatedspectra}{61\,313}
\newcommand{\repeatedobjects}{6\,094}
An estimate of some of the contributions to the internal errors associated with the analysis can be provided by injecting 
a large sample of interpolated (at random stellar parameter values)
noised synthetic spectra with different uncertainties in 
radial velocity %
\citep[see][Sec.~5]{worley_ambre_2012} into the pipeline. 
 We point out that we only test the performances of the \matisse \, method itself here
by checking its ability to retrieve the atmospheric
parameters in a very ideal case, since only a Gaussian white noise is assumed, and any possible
mismatch between the synthetic and real spectra are assumed to be negligible.
Since synthetic spectra that are almost the same spectral resolution and spectral coverage were used 
 for AMBRE:HARPS as for AMBRE:FEROS, the error analysis of AMBRE:FEROS 
can be considered as valid for the analysis of the AMBRE:HARPS spectra. As shown in \citet{worley_ambre_2012}:

\begin{itemize}
\item the behaviour of the 70$^{th}$ percentile of the internal error of each atmospheric parameter as a function of 
  the S/N show that for $\mathrm{S/N} > 10$ the internal errors are negligible and, thus, have almost no effect on the 
  determined parameters \citep[see][Fig.~10]{worley_ambre_2012}. Since only the HARPS spectra with $\mathrm{S/N} > 10$ 
  have been retained, this argument is valid for the present parameterisation.
\item The error on radial velocity for the accepted sample has a small effect on the determination of the stellar parameters. 
  We reached this conclusion on the basis of \citet[][Fig.~11]{worley_ambre_2012} and our Fig.~\ref{fig:vrad-err-hist}. 
    In \citet[][Fig.~11]{worley_ambre_2012}, the behaviour of the $70^\mathrm{th}$ percentile
    of the variation in each photospheric parameter calculated from the same spectrum as a function of artificial 
    variations in the value of $V_\mathrm{rad}$ is reported. There, it is shown that the 
    $70^\mathrm{th}$ percentile value for each parameter is relatively small for $\Delta V_\mathrm{rad} < 10$~km~s$^{-1}$ 
    (even smaller if $\Delta V_\mathrm{rad}$ is limited to 6~km~s$^{-1}$). Figure~\ref{fig:vrad-err-hist} shows that almost 
    all the spectra ($\simeq 99 \%$) that passed the rejection criteria described in Sec.~\ref{sec:rejection} have 
    a $\Delta V_\mathrm{rad} < 6$~km~s$^{-1}$. Putting together this result with what reported above, we conclude that
    we neglect hereafter the contributions of $V_\mathrm{rad}$ uncertainties and low-quality spectra to the internal 
    errors, since every spectra with $S/N<10$ and error in $V_\mathrm{rad}$ greater than 10 km~s$^{-1}$ has been rejected.
\end{itemize}

The other possible sources of internal errors (and particularly any possible
mismatch among observed spectra, real spectra and the effects of the real noise of the spectra)
have been investigated by considering a second independent method. 
It consisted in estimating the internal error of the AMBRE:HARPS analysis by comparing 
the parameterisation of the repeated observations of the same star, which are characterized by different 
S/N and uncertainty on the radial velocity.
 
For that purpose, the sample of \acceptedspectra\, parameterised spectra were investigated to identify those stars, 
which had been observed by HARPS at least 50 times, adopting a coordinate search radius of 5\arcsec. 
Using the web tool SIMBAD, we identified and then excluded from this repeat sample the observations of multiple stellar 
systems and variable stars, since the stellar parameters of these objects could vary with time.

This resulted in a sample of \repeatedspectra\, spectra that corresponded to \repeatedobjects\, distinct stars observed
50 times or more as per the radius search.
For each set of repeated observations, the mean value of each stellar parameter and the deviations 
from the mean values for each spectrum was calculated. In this way, we ended up with deviations from mean values as functions of S/N.
The sets of repeat spectra were then sorted by their S/N to compute the 0.7 quantile of the parameter deviations 
for bins of $\Delta\mathrm{S/N} = 20$.

\begin{figure}
  \centering
  \resizebox{\hsize}{!}{\includegraphics{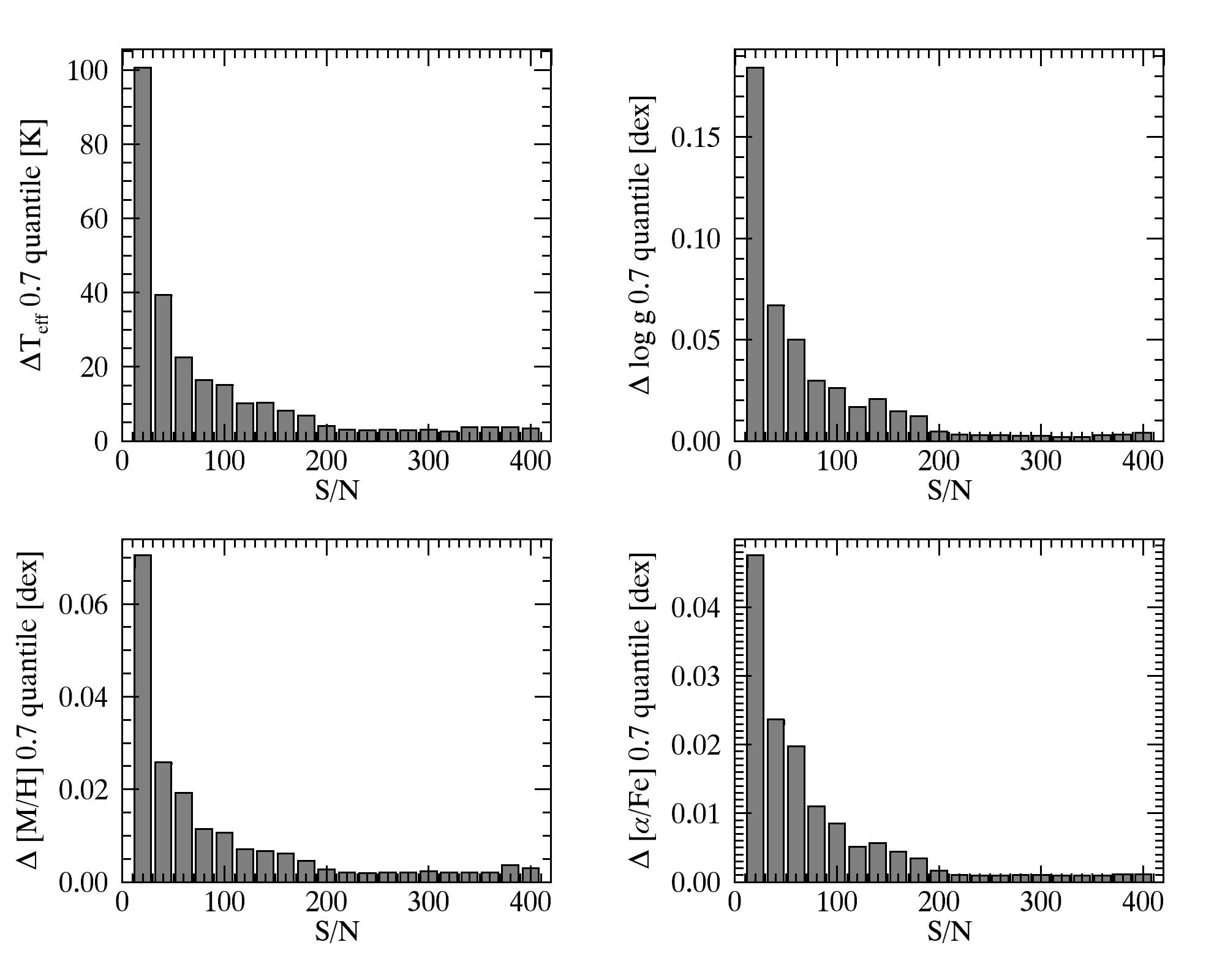}}
  \caption{Histograms of the changes for each atmospheric parameter as a function of the S/N for the repeated spectra. 
      For each bin, the 0.7 quantile of the deviation from the mean value is shown. These 0.7 quantiles define 
      the internal errors associated with each parameter.}
  \label{fig:repeat-error}
\end{figure}

The results reported in Fig.~\ref{fig:repeat-error} show that the deviations in each stellar parameter decrease with 
increasing S/N, as expected. These trends were used to define the internal errors as follows. For a given spectrum with 
an associated S/N, the 0.7 quantile-value in the corresponding S/N-bin was adopted as the estimate of the internal error 
of the AMBRE:HARPS parameterisation.

For spectra with S/N > 160, a lower limit for the internal errors (set by the MATISSE internal error) was adopted as follows:
\begin{itemize}
\item $\sigma_\mathrm{int}(T_\mathrm{eff}) = 10$ K;
\item $\sigma_\mathrm{int}(\log g) = 0.02$ dex;
\item $\sigma_\mathrm{int}(\mathrm{[M/H]}) = 0.01$ dex;
\item $\sigma_\mathrm{int}([\alpha/\mathrm{Fe}]) = 0.005$ dex.
\end{itemize}

\section{External error analysis}
\label{sec:external-err}
\newcommand{\gestbias}{$\simeq-67$~K}
\newcommand{\geststd}{$\sim$31~K}
\newcommand{\gesloggbias}{$-0.13$~dex}
\newcommand{\gesloggstd}{$\simeq0.06$~dex}
\newcommand{\gesmbias}{0.05~dex}
\newcommand{\gesmstd}{0.06~dex}

To quantify the external errors associated with the AMBRE:HARPS stellar parameters, they were compared with literature values
for key samples within the dataset.

\subsection{Benchmark stars}
\label{sec:reference-samp}
\begin{figure}[tb]
  \centering
  \subfloat%
  {\includegraphics[scale=.4]{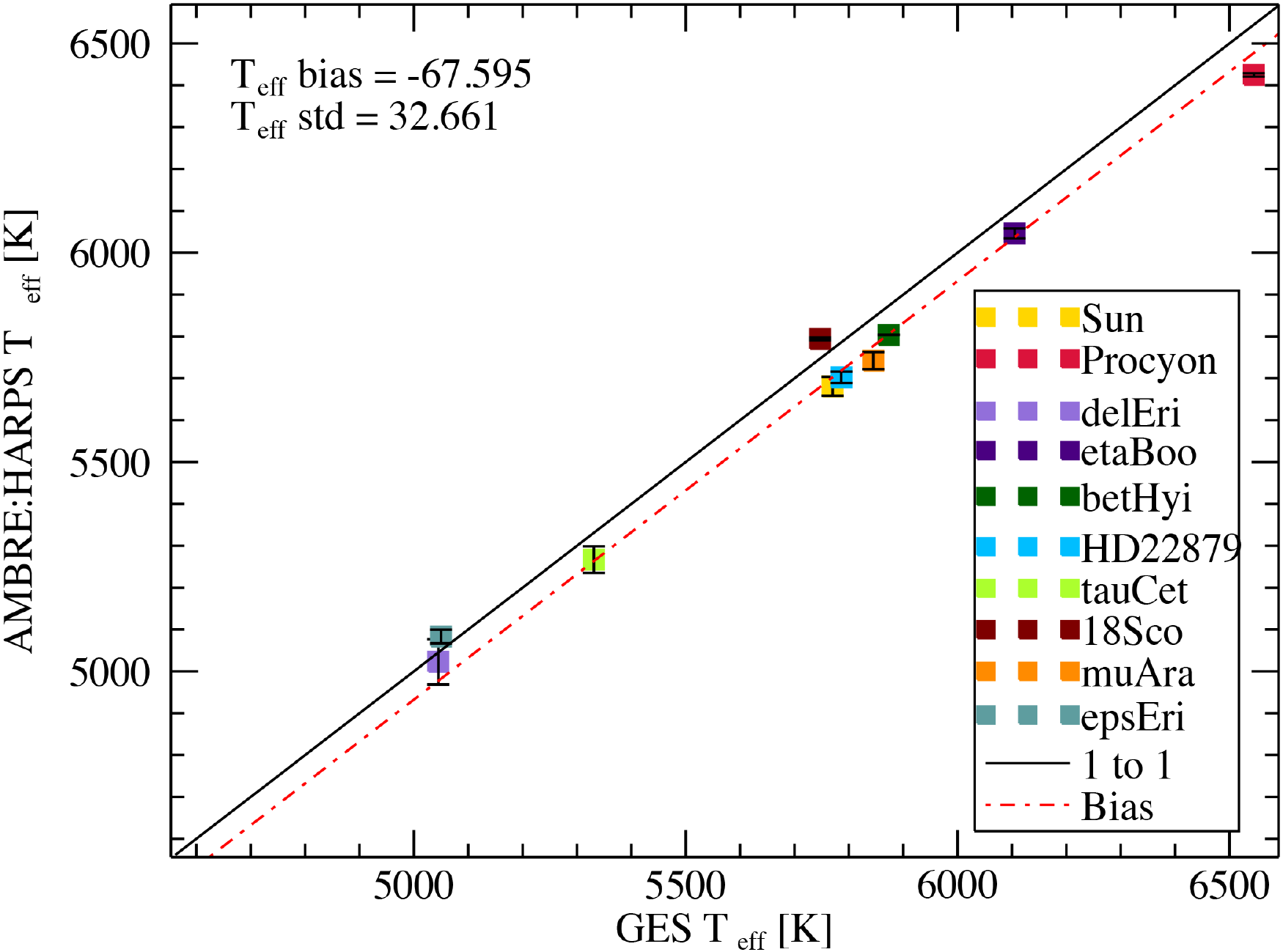}} \\
  \subfloat%
  {\includegraphics[scale=.4]{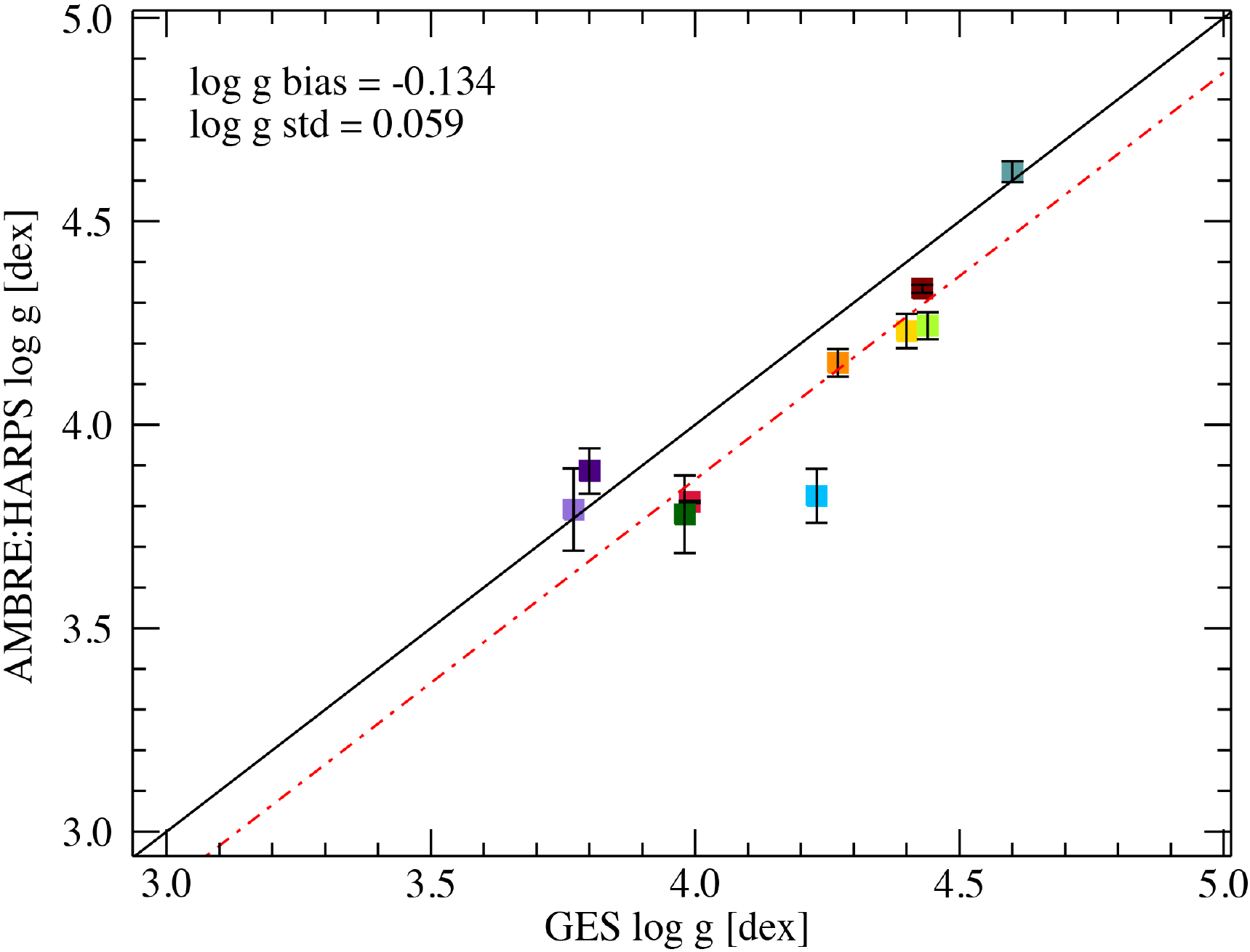}} \\
  \subfloat%
  {\includegraphics[scale=.4]{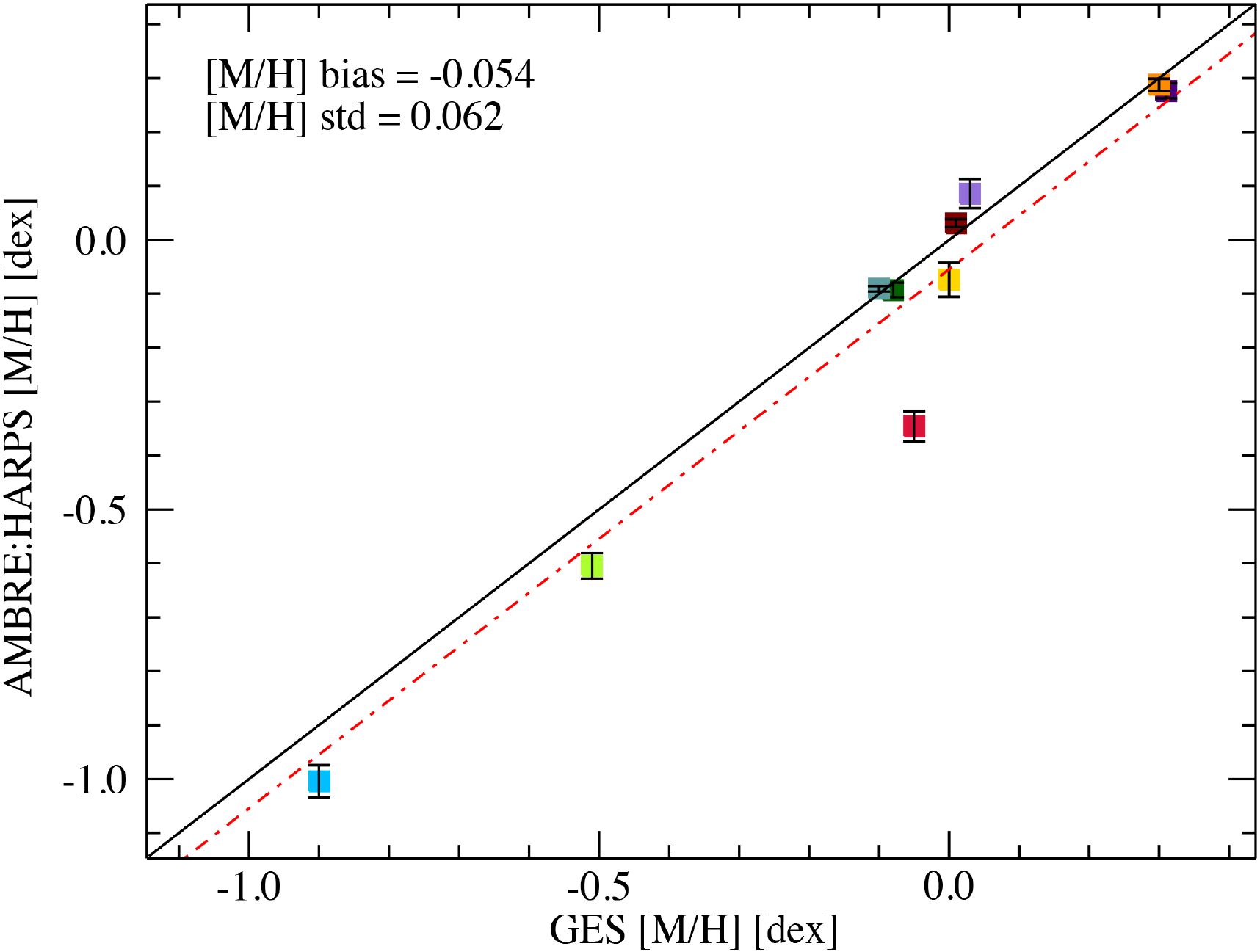}} 
  \caption{Comparison between the star atmospheric parameters for AMBRE:HARPS and FGK benchmarks from 
    \citet[][]{jofre_fgk_2014} and Heiter et al. (2014, in prep.).
    The comparison is performed for $T_\mathrm{eff}$ (top panel), $\log g$ (middle panel), 
    and [M/H] (lower panel). Each point in the different panels corresponds to the mean value 
    of the corresponding parameter when several AMBRE:HARPS spectra are available. 
    For each parameter, the vertical error bars represent the standard deviation from the mean 
    AMBRE:HARPS value.}
  \label{fig:ges-vs-harps}
\end{figure}

As a first estimate of the quality of this analysis, the results of the AMBRE:HARPS pipeline were investigated for a sample 
of well-studied reference stars. This sample is the FGK benchmark star sample defined for the Gaia mission and the Gaia ESO Survey 
(GES), for which the parameters are based on the homogeneous analysis of high quality data.
For these benchmarks, we adopted the effective temperatures and surface gravities of 
Heiter et al. (2014, in prep.) with the metallicities from \citet[][]{jofre_fgk_2014}. 
Currently, there are no reference abundances of the $\alpha$-elements available for these stars.
We show the comparison between the AMBRE:HARPS and reference values 
for the stellar parameters of the ten FGK benchmark stars, which have a HARPS spectrum with 
S/N~$> 60$ within the AMBRE:HARPS dataset in the three panels of Fig.~\ref{fig:ges-vs-harps}. 
The agreement is very good between both set of parameters with low biases and standard deviations, 
validating the results of the AMBRE:HARPS analysis for cool dwarfs with metallicities higher than $-1.0$~dex (i.e. 
the greater majority of the AMBRE:HARPS sample, see below). We, however, point out that the agreement is slightly less
good for the derived surface gravity of HD~22879 and the overall metallicity of the hot star Procyon, whose
spectrum exhibits less metallic lines than the bulk of the AMBRE:HARPS sample.

\subsection{Porto sample}
\label{sec:porto-samp}
\newcommand{\tbias}{$-59$~K}
\newcommand{\tstd}{$\sim$87~K}
\newcommand{\loggbias}{$-0.14$~dex}
\newcommand{\loggstd}{0.19~dex}
\newcommand{\mbias}{null}
\newcommand{\mstd}{0.1~dex}
\newcommand{\alphabias}{null}
\newcommand{\alphastd}{0.03~dex}
\newcommand{\matchingstars}{713}
\newcommand{\matchingspectra}{3\,991}
An independent group previously published a large dataset of stellar atmospheric parameters estimated from HARPS spectra 
(Porto sample, hereafter) \cite[see][]{sousa_spectroscopic_2008,sousa_spectroscopic_2011,sousa_spectroscopic_2011-1,adibekyan_chemical_2012,tsantaki_deriving_2013}. This sample 
consists in $1\,111$ stars for which stellar parameters have been estimated using a completely different method than that 
employed here, since the Porto method is based on an equivalent widths analysis of several selected lines.

The results of the Porto analysis provide a unique opportunity to perform a comparison between the parameterisation performed 
by two independent methods (with different line lists, model atmospheres, spectral analysis techniques, etc.) that analysed 
the same significantly large quantity of high quality spectra produced by the same instrument.
Furthermore, this allows us to conduct a robust estimation of the external errors of the AMBRE:HARPS pipeline. Before describing the comparison between the two samples, 
we remind here that the AMBRE pipeline derives a mean [M/H] (i.e. taking into account all the metals),
while the Porto group derives the [Fe/H] metallicity by considering only iron lines. 

\begin{figure}[h]
  \centering
  \resizebox{\hsize}{!}{\includegraphics{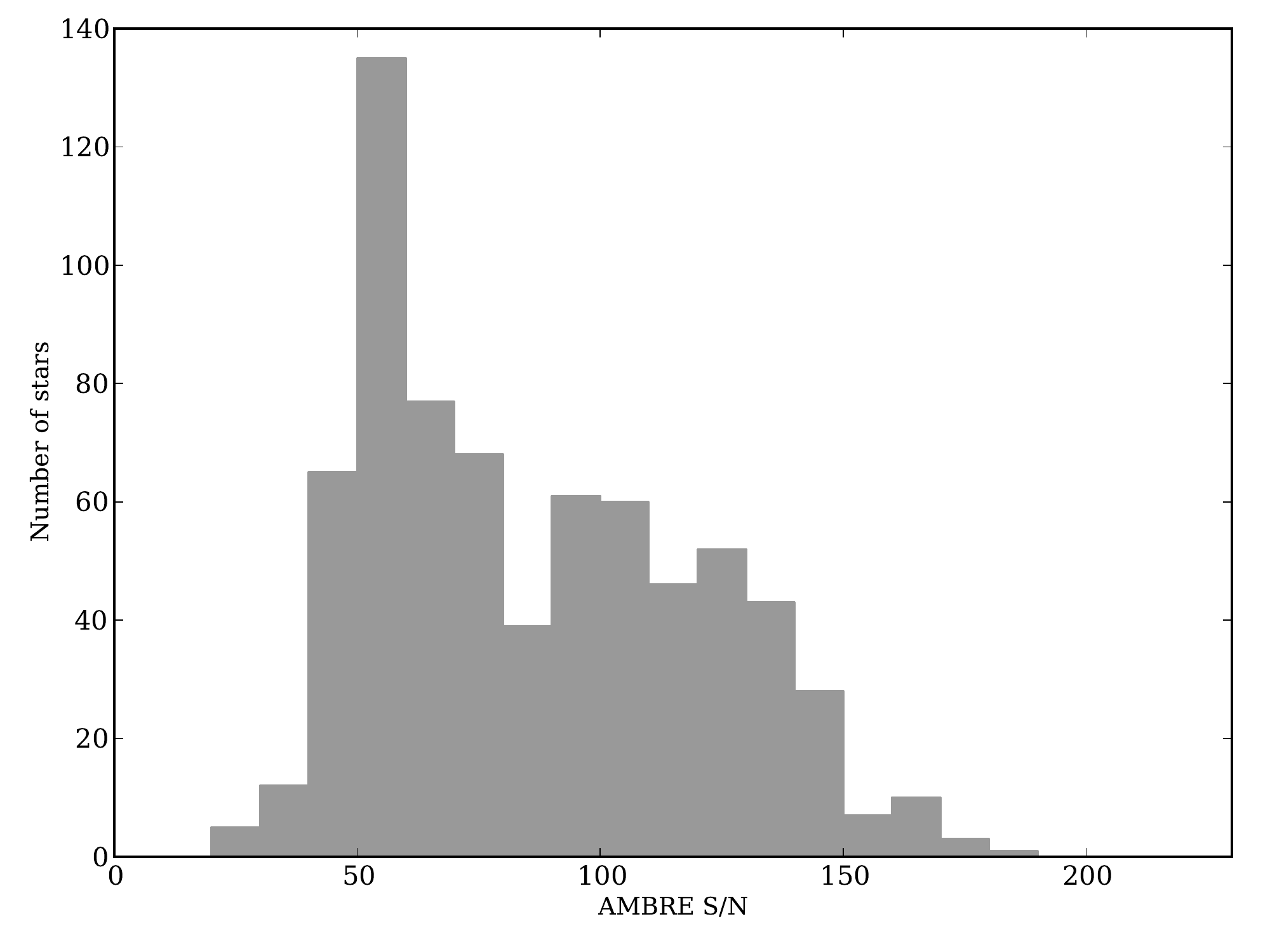}}
  \caption{Histogram of S/N for stars in the AMBRE:HARPS dataset matching the Porto sample.}
  \label{fig:ambre-snr-hist}
\end{figure}

\begin{figure*}[h!]
  \centering
  \includegraphics[width=17cm]{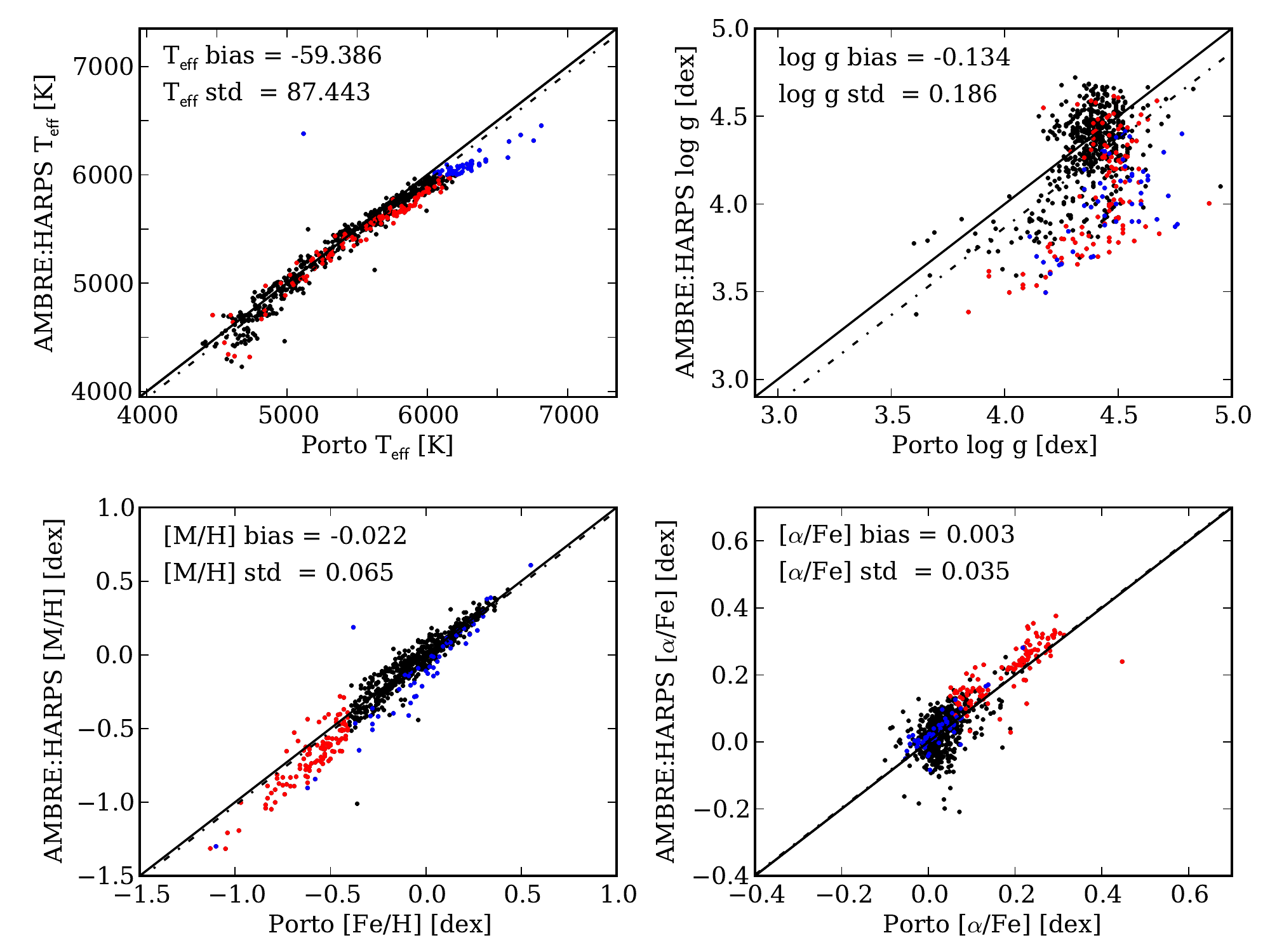}
  \caption{Comparison between the stellar atmospheric parameters derived by the AMBRE:HARPS pipeline and the reference sample from Porto. 
    The solid line in each panel traces the one-to-one relation, while the dot-dashed line shows the location of the bias between the samples. 
    Stars with $T_\mathrm{eff} > 6\,000$ K are
    plotted in the panels with blue dots; stars with $\textrm{[Fe/H]} < -0.4$ dex are the red dots.
     The biases in $T_\mathrm{eff}$, $\log g$, [M/H] and [$\alpha$/Fe] for stars marked with red dots 
    are -90.1 K, -0.349 dex, -0.098 dex, and 0.030 dex, respectively. For stars marked with blue dots the biases 
    are -161.3 K, -0.451 dex, -0.093 dex, and 0.011 dex, respectively.}
  \label{fig:porto-vs-ambre}
\end{figure*}

A cross match between the coordinates of the Porto sample and AMBRE:HARPS using a radius of $\sim$5~\arcsec revealed \matchingstars\, 
stars have been analysed by both methods. These \matchingstars\, stars correspond to \matchingspectra\, spectra in the 
AMBRE:HARPS dataset with S/N ranging from $\sim$20 to $\sim$221 (see Fig.~\ref{fig:ambre-snr-hist}). For the purpose of this comparison, the mean value of 
the derived AMBRE:HARPS stellar parameters (and the 
associated standard deviations) were calculated to represent the star when several spectra were available for the same star. 
The AMBRE:HARPS parameters were then compared to the Porto values in Fig.~\ref{fig:porto-vs-ambre}. 

\begin{figure*}[h]
  \centering
  \includegraphics[width=17cm]{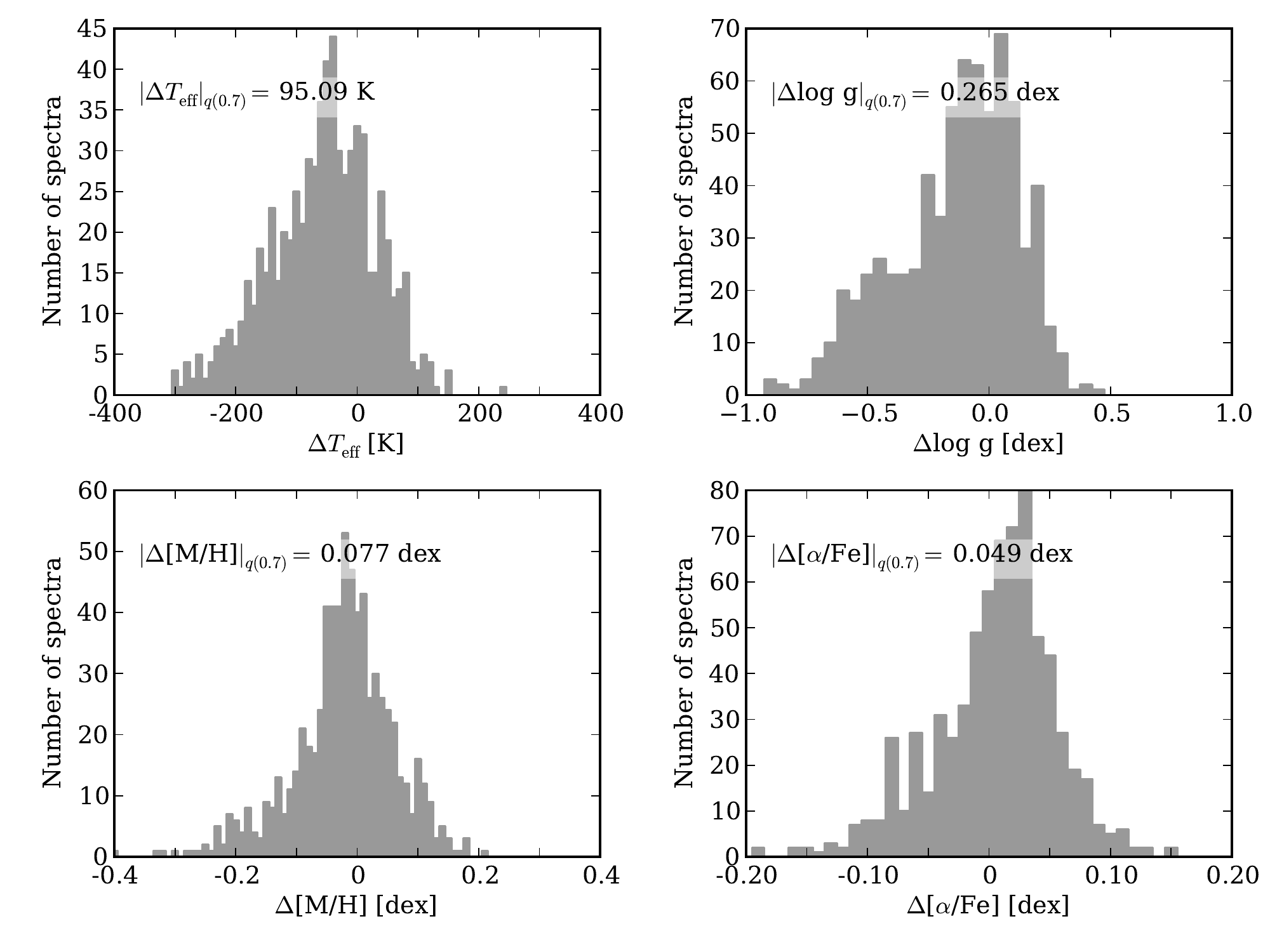}
  \caption{Distribution of the residuals $\theta_i^\textrm{AMBRE} - \theta_i^\textrm{Porto}$ for the stellar parameters 
    $\theta_i$ = ($T_\mathrm{eff}$, $\log g$, [M/H] and [$\alpha$/Fe]). For the AMBRE:HARPS stars with repeated observations, 
    the mean of the AMBRE:HARPS derived parameters are shown. The asymmetries in the distributions reflect the 
      asymmetries to the 1 to 1 lines that can be seen in Fig.~\ref{fig:porto-vs-ambre}.
    In each panel the 0.7 
    quantile of the absolute value of the residuals corrected for the 
    bias is reported.}
  \label{fig:porto-ambre-errors}
\end{figure*}

The agreement between the Porto and AMBRE:HARPS stellar parameters are also illustrated in Fig.~\ref{fig:porto-ambre-errors} where 
the distribution of the residuals from the median value for each parameters are shown. The main characteristic of these
 distributions is that they are not perfectly centered on zero, but rather small biases exist that have been estimated as the mean 
value of the differences between Porto and AMBRE:HARPS. 

Figure~\ref{fig:porto-vs-ambre}-top left shows the comparison between the two set of effective temperatures. Despite the difference 
in the two methods, there is very good agreement between both results, the bias being around 
\tbias, and the dispersion around \tstd. It can be noted that the updated Porto effective temperatures of \citet[][]{tsantaki_deriving_2013} 
for the cooler stars (below 5\,000~K) are based on a new line list specifically built for these cool stars, which considerably 
improved the agreement in that temperature range. Indeed, their previous effective temperatures were slightly higher, leading to 
an absolute value of the bias between Porto and AMBRE:HARPS of 86~K (with standard deviation of 128 K) for the effective temperatures 
between 4\,000 and 5\,000~K. These bias and standard deviations decreased to an absolute value of 26~K and 117 K, respectively, 
when adopting the Porto revised values. In Figure~\ref{fig:porto-vs-ambre} we also point out that, 
the small tail at large negative values is produced by stars hotter than 6\,000~K, which are more difficult to parametrize
than cooler stars.

The agreement is not as good but still very reasonable for the stellar gravity comparison (see Fig.~\ref{fig:porto-vs-ambre}), 
as it is probably the most difficult stellar atmospheric parameter to derive. The bias in gravity is \loggbias, and the standard 
deviation is \loggstd. In Fig.~\ref{fig:porto-vs-ambre}, it can be seen that the possible cause of this bias and standard 
deviation could be that the Porto stellar gravities seem to span a smaller range of values than those obtained with the AMBRE:HARPS pipeline. 
Most of their sample data are indeed found between 4.3 and 4.6~dex, while the AMBRE:HARPS are mostly found between 4.2 and 4.7~dex. 
It can be seen in this figure that the large majority of the discrepant stars have an effective temperature
larger than 6\,000~K and/or a metallicity smaller than -0.4~dex, i.e. the cases for which spectra exhibit rather few lines to 
perform the parameterisation.

On the other hand, Fig.~\ref{fig:porto-vs-ambre} reveals that the agreement between the stellar metallicities is very good
(the bias is almost \mbias\, and the dispersion is smaller than \mstd). We note that we derive a slightly lower value than Porto, 
for stars with a mean metallicity smaller than -0.4~dex, and the 
small discrepancy seems to increase for lower metallicities. Indeed, when considering only stars, which have [M/H]<-0.4~dex, we calculate 
a bias and a standard deviation of -0.091~dex and 0.076~dex, respectively. Moreover, we point out that these stars 
are $\alpha$-enriched (see Fig.~\ref{fig:porto-vs-ambre}). 
In any case, the agreement is much better for more metal-rich stars, which consists of the bulk of the total sample. 

\begin{figure}[h]
  \centering
  \resizebox{\hsize}{!}{\includegraphics{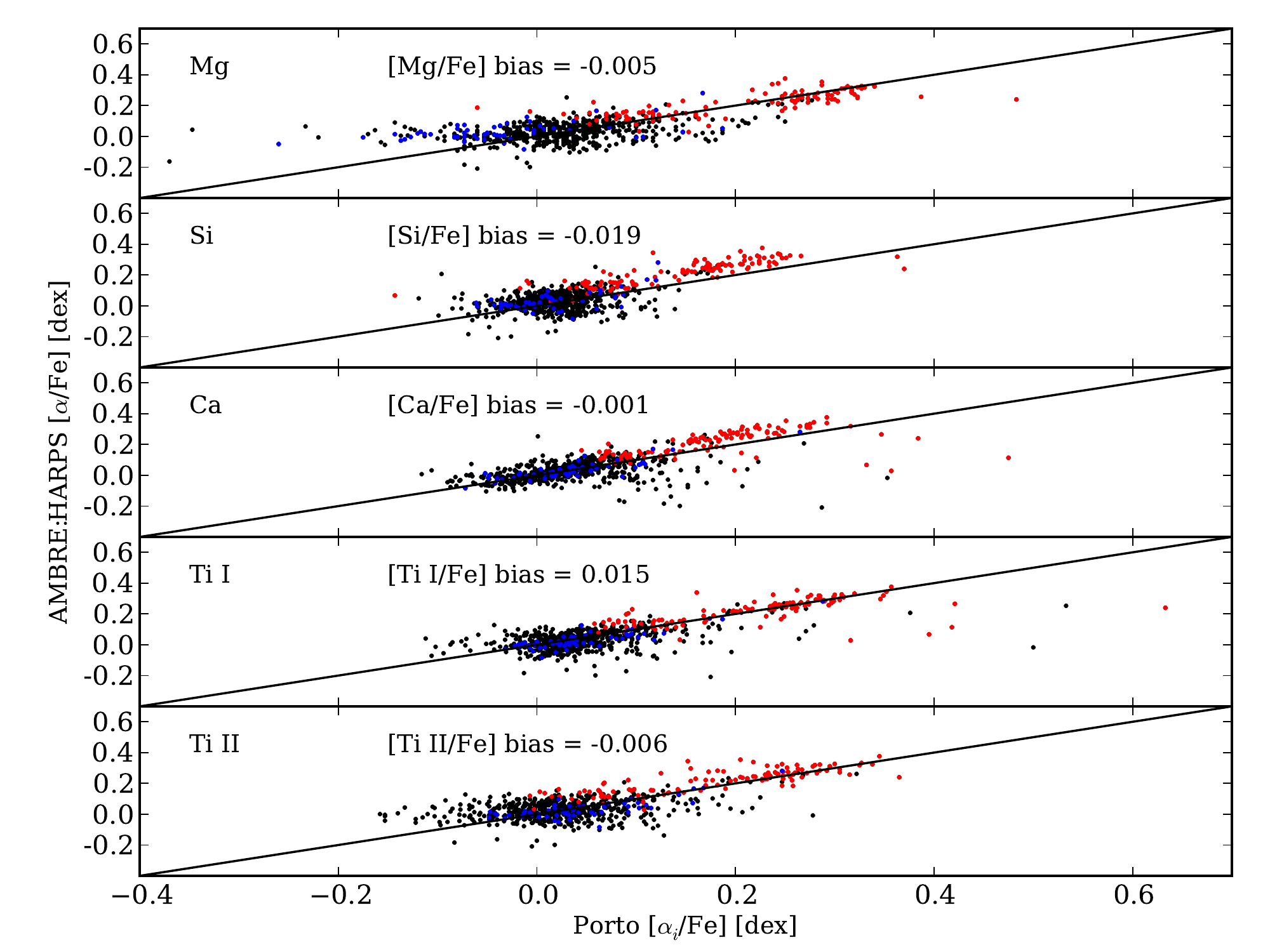}}
  \caption{ Comparison between the mean $\alpha$-abundances derived by the AMBRE pipeline and the abundances of the 
    individual     
    $\alpha$-elements determined by Porto. The largest departures from the 
    1-to-1 line are found for the \ion{Si} \ and \ion{Ca} \ abundances and could explain the 
    departure observed  
    in Fig.~\ref{fig:porto-vs-ambre} for the most metal-poor stars.   
    As in Fig.~\ref{fig:porto-ambre-errors}, stars with $T_\mathrm{eff} > 6\,000$ K are plotted in the panels with blue dots; stars with 
    $\textrm{[\ion{Fe}/\ion{H}]} < -0.4$ dex are the red dots.}
  \label{fig:alpha-porto}
\end{figure}

Finally, it was also possible to compare the abundance ratios of the $\alpha$-elements with respect to iron. \citet{adibekyan_chemical_2012} 
published the individual abundances of the following $\alpha$-species: \ion{Mg}, \ion{Si}, \ion{Ca}, \ion{Ti}{I}, and \ion{Ti}{II}. The mean of these five abundances was 
taken to compare to the AMBRE:HARPS [$\alpha$/Fe] ratios (see Fig.~\ref{fig:porto-vs-ambre}-lower right panel). Once again, the 
agreement is very satisfactory with a quasi-\alphabias\, bias and a standard deviation of only \alphastd. It can be, however, noted that, 
 a small departure for this [$\alpha$/Fe] ratio comparison, from the one-to-one line is present for the highest values of [$\alpha$/Fe], or 
the most metal-poor stars (the Porto sample having slightly smaller [$\alpha$/Fe] ratios when [$\alpha$/Fe] $>$0.2~dex). To understand this behaviour, 
we studied the values of the mean $\alpha$ abundances produced by our pipeline as a function of the abundances of the individual 
$\alpha$-species provided by \citet{adibekyan_chemical_2012}. Figure~\ref{fig:alpha-porto} reveals that the disagreement is probably 
mostly caused by the behaviour of the \ion{Ca}  \ and \ion{Si}  \ abundances that depart more than the other $\alpha$-species.

In summary, it can be concluded that the agreement between the Porto and AMBRE:HARPS stellar parameters are very satisfactory. This comparison sample 
consists mainly of cool dwarf (solar-type) stars, and the agreement, therefore, validates the AMBRE:HARPS results for more than 90\% of the spectra 
found in the AMBRE:HARPS sample. It has, however, to be noted that a lack of a good comparison sample did not allow us to estimate the systematic 
differences for non solar-type stars (although some estimates are given in Fig.~\ref{fig:porto-vs-ambre} for hot and
metal-poor dwarfs, respectively). Finally, we point out that the Porto sample is also probably affected by trends and systematics and
that some of the differences with AMBRE:HARPS could also partly originate in the Porto catalogue.

\subsection{External error quantification}
\label{sec:extern-error-quant}
 From the above comparisons, we can conclude that only small biases are found between the AMBRE:HARPS stellar parameters and 
those of the Gaia Benchmark and Porto samples. We therefore decide to not perform any bias corrections, contrary to what has been 
done for the AMBRE:FEROS parameters. In this way, we also avoid needing to correct the stellar parameters of the remainder of 
AMBRE:HARPS spectra sample that are not found within the reference and Porto samples.

As for the external errors associated with the AMBRE:HARPS parameters, they have been defined using the Porto sample, which is a 
statistically significant comparison sample. It must be noted that the Porto sample has its own sources of error, and therefore, 
the adopted external errors are probably overestimated. These external errors are estimated from the 0.7 quantiles reported in 
Fig.~\ref{fig:porto-ambre-errors} and are
\begin{itemize}
\item $\sigma_{T_\mathrm{eff},\mathrm{ext}} = 93~$ K;
\item $\sigma_{\log g,\mathrm{ext}} = 0.26$ dex;
\item $\sigma_{\mathrm{[M/H]},\mathrm{ext}} = 0.08$ dex;
\item $\sigma_{[\alpha/\mathrm{Fe}],\mathrm{ext}} = 0.04$ dex.
\end{itemize}
We point out that these external errors are based on dwarf star comparisons only. However, these values were also adopted 
for the few giants of the AMBRE:HARPS sample since (i) there was a lack of a good reference sample for giants in
AMBRE:HARPS and (ii) as shown in \citet{worley_ambre_2012}, except perhaps for the surface gravity, the external errors 
can be considered as constant for most of the stellar types.

As the final error associated to each parameter will be the quadratic sum of the internal and external errors, 
we verified the consistency of this value with the deviations of the AMBRE:HAPRS parameters from the reference.
Referring to the Porto sample, we confirmed that the error associated to AMBRE:HARPS parameters is a good
approximation to the deviations from the literature at all S/N.

\section{The AMBRE:HARPS stellar parameters delivered to ESO}
\label{sec:results}

\subsection{Final parameters of AMBRE:HARPS analysis}
\label{sec:final-param}
\begin{figure}[h!]
  \centering
  \resizebox{\hsize}{!}{\includegraphics{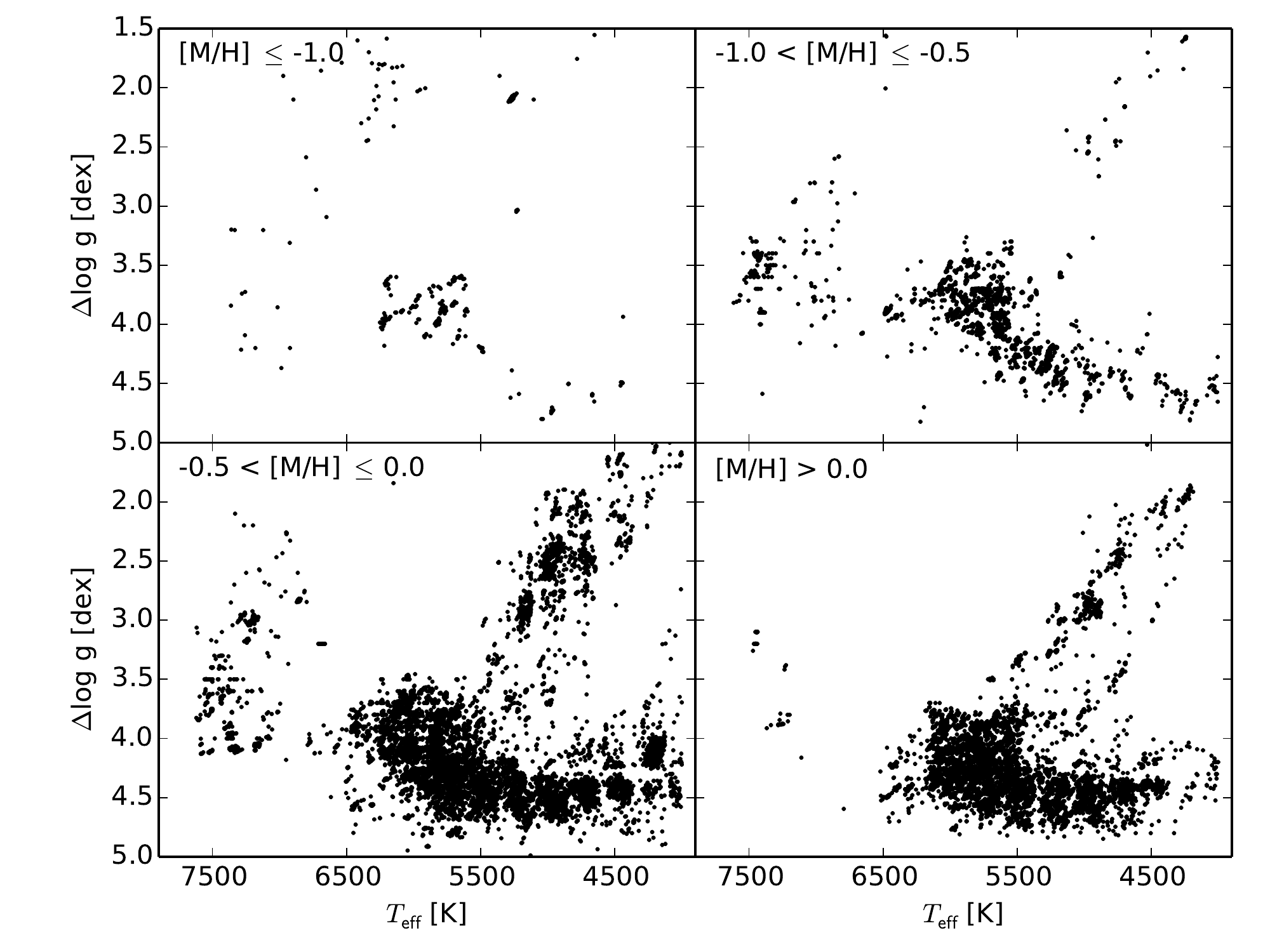}}
  \caption{ HR diagram of the AMBRE:HARPS stellar atmospheric parameters. The different panels
    correspond to different 
    [M/H] bins, as indicated in the figure. Almost 14\% of the retained spectra have a metallicity 
    [M/H]$< -0.5$ dex.}
  \label{fig:hr-diag-spc}
\end{figure}

\begin{figure}
  \centering
  \resizebox{\hsize}{!}{\includegraphics{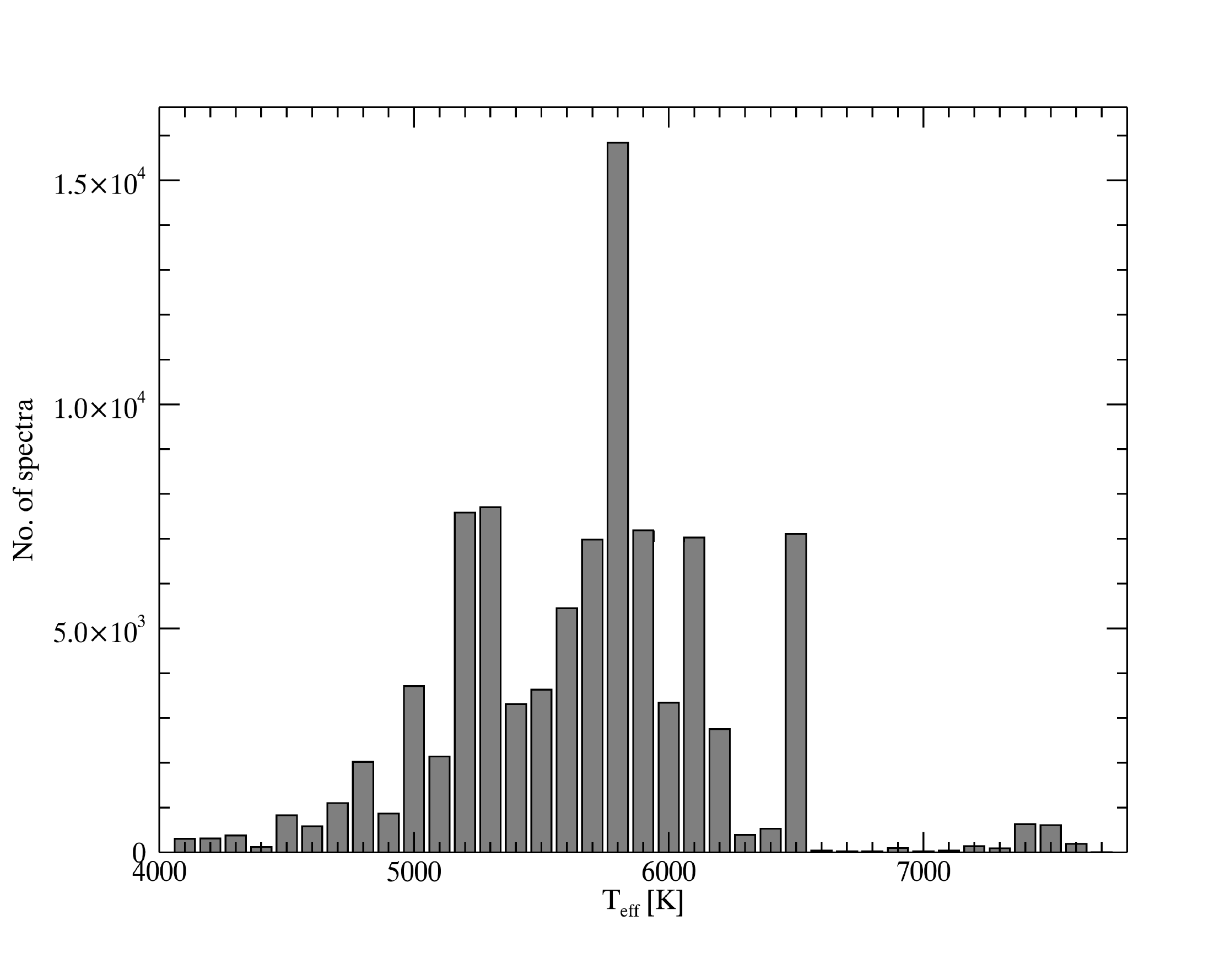}}
  \caption{Distribution of the derived AMBRE:HARPS $T_\textrm{eff}$ values. The distribution has a main peak around 
    5\,700~K and a second one between 5200~K and 5300~K.}
  \label{fig:t-eff-hist}
\end{figure}

\begin{figure}
  \centering
  \resizebox{\hsize}{!}{\includegraphics{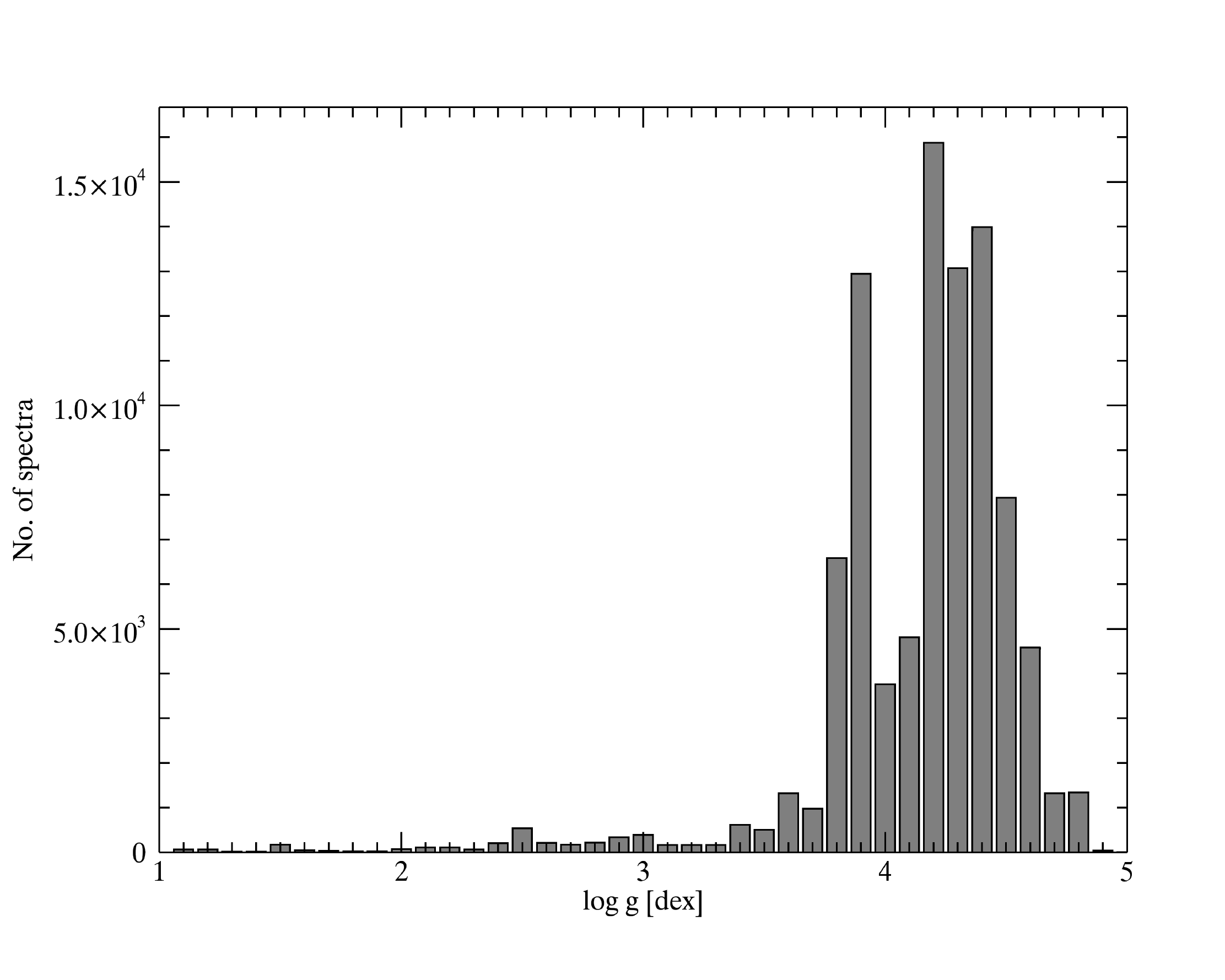}}
  \caption{Distribution of the derived AMBRE:HARPS $\log g$ values. This distribution peaks around the
solar gravity.}
  \label{fig:logg-hist}
\end{figure}

\begin{figure}
  \centering
  \resizebox{\hsize}{!}{\includegraphics{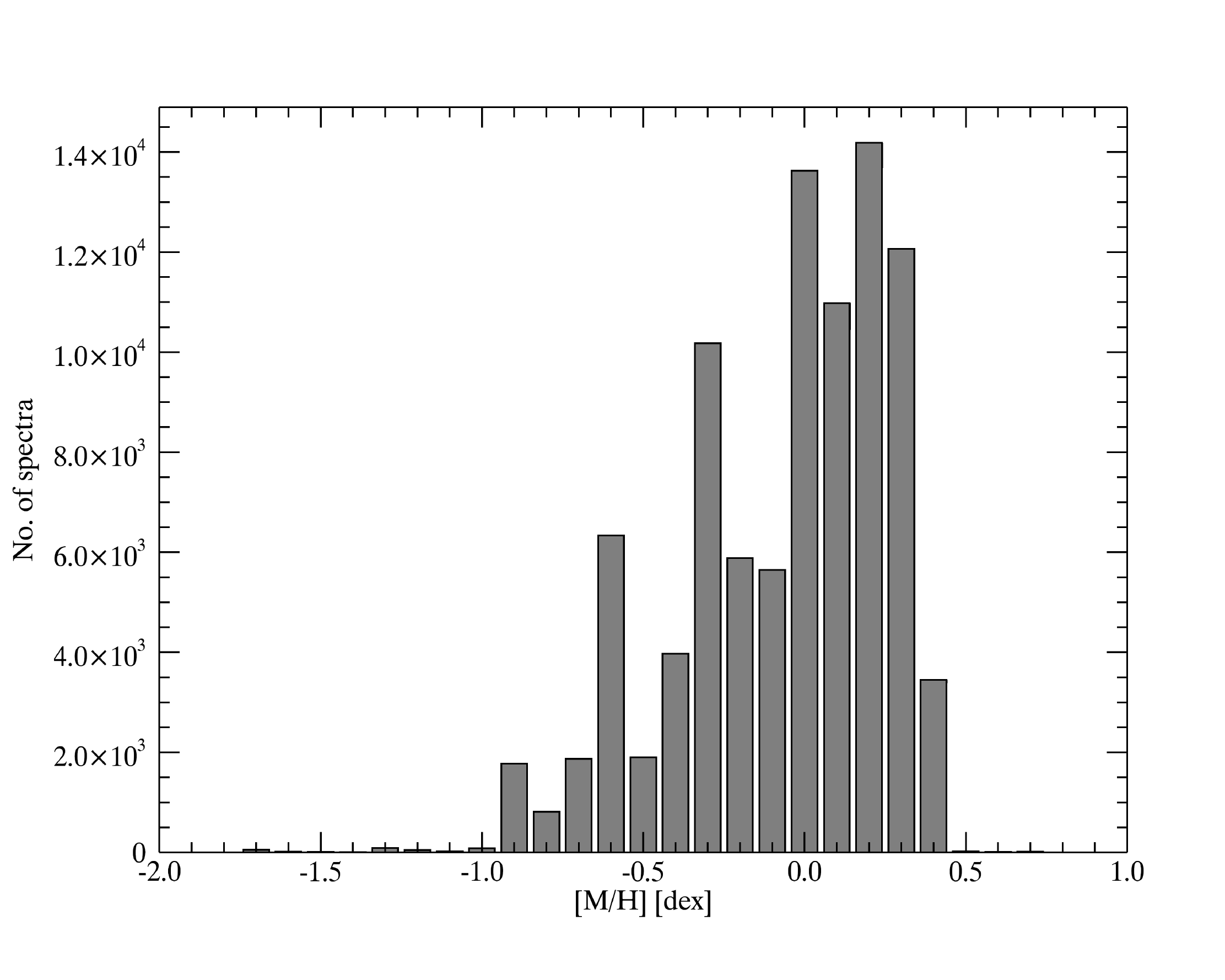}}
  \caption{Distribution of the derived AMBRE:HARPS metallicities.}
  \label{fig:m-over-h-hist}
\end{figure}

\begin{figure*}[h]
  \centering
  \includegraphics[width=17cm]{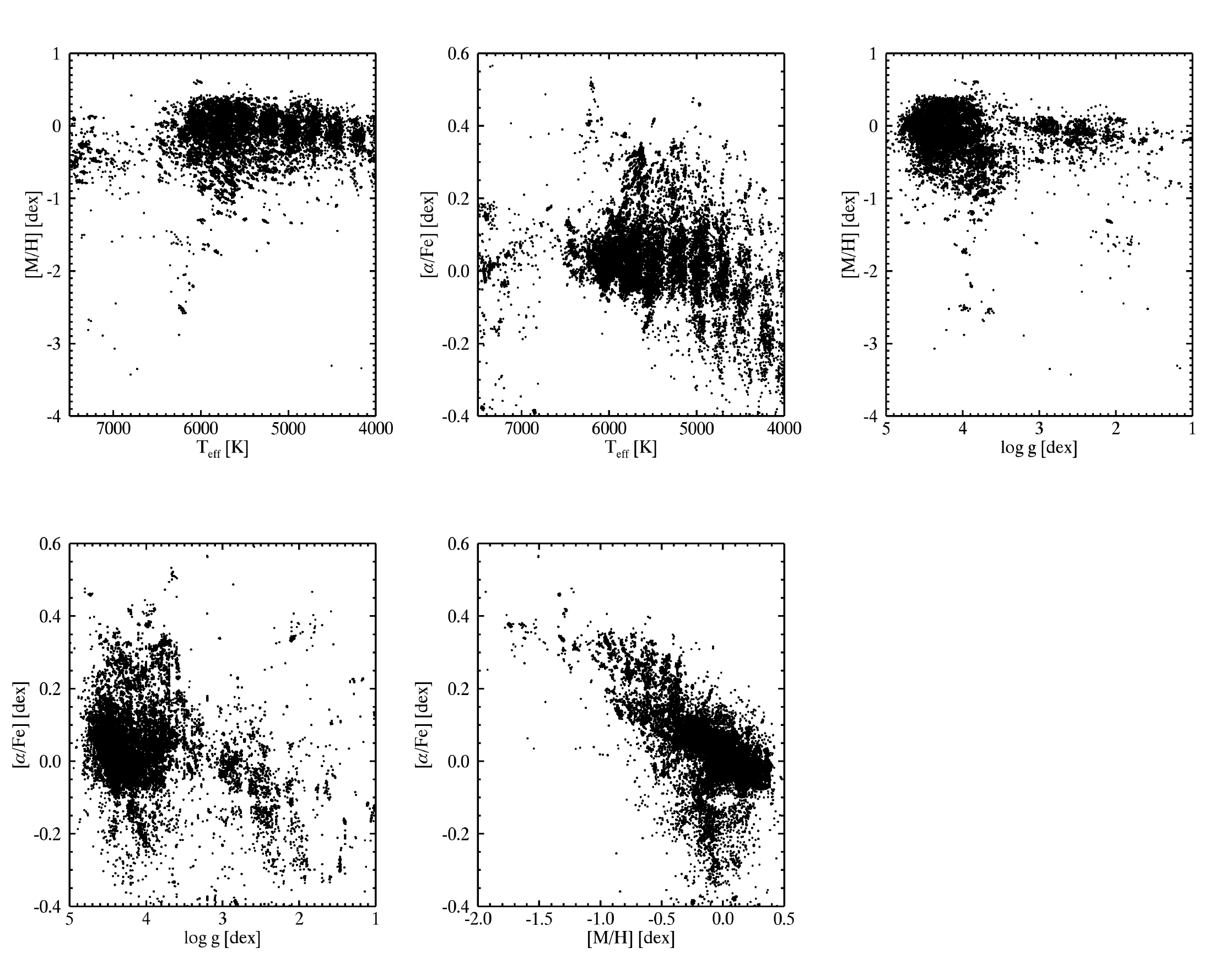}
  \caption{The final AMBRE:HARPS stellar atmospheric parameters. Different combinations of the four
    derived parameters. Top-left: [M/H] vs. $T_{\mathrm{eff}}$. Top-center: [$\alpha$/Fe] vs. $T_{\mathrm{eff}}$. 
    Top-right: [M/H] vs. $\log g$. Lower-left: [$\alpha$/Fe] vs. $\log g$ and Lower-center: 
    [M/H] vs. [$\alpha$/Fe].}
 \label{fig:final_harps}
\end{figure*}

The final accepted stellar parameters of the \acceptedspectra\, AMBRE:HARPS spectra are shown in Fig.~\ref{fig:hr-diag-spc} 
(HR diagram), Fig.~\ref{fig:t-eff-hist}-~\ref{fig:m-over-h-hist} (histograms of the distribution of the three main 
stellar parameters), and Fig.~\ref{fig:final_harps} (different combinations of the AMBRE:HARPS stellar parameters). 

The main characteristic of Fig.~\ref{fig:hr-diag-spc} is that the great majority of the spectra are populating the 
main sequence, confirming that more than 90\% of the entire HARPS:AMBRE sample constitutes of cool dwarf stars of 
G and K spectral types (see also Fig.~\ref{fig:t-eff-hist}, \ref{fig:logg-hist}, and Fig.~\ref{fig:final_harps}).
Actually, a large fraction ($\simeq 42 \%$) of the spectra correspond to stars with an effective temperature
close to the solar. Furthermore, most of these dwarfs ($\sim$86\%) have a solar metallicity, which larger than -0.5~dex 
(see Fig.~\ref{fig:m-over-h-hist}). The red giant branch is clearly seen, although it is much less populated. From 
Fig.~\ref{fig:logg-hist}, it can be seen that about 4\% of the total sample is composed of giant stars 
(defined as $\log g < 3.5$ dex). We refer to a forthcoming paper for a deeper analysis of the AMBRE:HARPS sample characteristics.

We finally point out that MATISSE relies on a learning phase based on the discrete grid of synthetic 
spectra. The products are vectors on which the observed spectra are projected to retrieve their 
parameters. For the AMBRE analysis, we have decided to adopt a version of these projection vectors computed 
from a direct inversion of the correlation matrix of the synthetic spectra \citep[see][]{kordopatis2011}. 
This assumption, giving better results for high quality spectra, can lead in some cases to pixelization 
effects due to an overfitting of the data (as it can be noticed in Fig~\ref{fig:final_harps}).

\subsection{ESO table description}
\label{sec:eso-table-descr}
The derived stellar parameters and [$\alpha$/Fe] abundances of the 90\,174 AMBRE:HARPS spectra are being ingested into 
the ESO archives (see http://archive.eso.org/cms/eso-archive-news/first-data-release-from-the-matisse-oca-eso-project-ambre.html). 
Table~\ref{tab:param-desc} describes the exact data that is being delivered to ESO.

To the values of the parameters calculated by the AMBRE:HARPS pipeline, the internal and external errors for each of the four 
parameters as defined in Sec.~\ref{sec:internal-err} and Sec.~\ref{sec:external-err}, respectively, have been added. The null 
value adopted for each parameter is also reported in this table. Moreover, each spectra has been flagged as a function of the 
$\chi^2$, which is derived from the agreement between the synthetic and the observed spectrum. As reported in Table~\ref{tab:param-desc}, 
three values have been adopted for this flag: {\it very good}, {\it good}, and {\it acceptable}. 

The limits applied by which to attribute these flags are defined differently for each temperature domain using as reference the interpolation 
curves that describe the $\chi^2$ as a function of the S/N (see Sec.~\ref{sec:rejection}). The percentage of spectra associated with 
each flag are reported in Table~\ref{tab:chi2-flag} with the limiting values on the $\chi^2$ defining the flag itself. The 
great majority of spectra that have a \textit{good} or \textit{very good} $\chi^2$ are observations from cold or warm stars ($4\,000 \mathrm{K} \leq 
T_\mathrm{eff} \leq 6\,500 \mathrm{K}$). On the other hand, 60\% of the spectra of the hot stars ($T_\mathrm{eff} > 6500 \mathrm{K}$) 
show a worse $\chi^2$. These results reflect the ability of any spectral analysis method (including MATISSE) to better 
determine reliable parameters for stars with effective temperatures between $\sim$4\,000~K and $\sim$7\,000~K (namely F, G and K-type 
stars). For hotter stars (or metal-poor ones), fewer spectral signatures are available leading to more uncertainty in the derived 
stellar parameters.

\section{Summary}
\label{sec:summary}
We have presented the automatic stellar parameter determination of more than 90\,000 HARPS spectra collected between 2003 and 2010 and archived at ESO. 
These spectra correspond to more than 10\,000 different stars that have been observed between one and several hundreds of time.

Stellar parameters have been determined for more than 70\% of the total sample of HARPS spectra delivered by ESO. The main rejection 
criteria are possibly from low S/N, broad line (e.g. due to fast-rotation) spectra, 
 poor synthetic fit of the observed spectra, and parameters found 
outside the synthetic spectra grid boundaries. The stellar parameters obtained for the HARPS spectra, which were derived using the 
AMBRE pipeline based on the MATISSE algorithm, are being delivered to ESO for ingestion in their archives.

The stellar parameters are the effective temperature, the surface gravity, the mean metallicity ([M/H]), the abundance of the 
$\alpha$-elements with respect to iron ([$\alpha$/Fe]), and their associated internal and external errors. We also provide the 
radial velocity (with the associated error and the FWHM of the corresponding CCF) and a quality flag for the parameters.

The AMBRE:HARPS sample is an extensive dataset of mainly solar-like stars for which the stellar parameters have been homogeneously 
determined. Given the richness of the HARPS archival data, the present dataset constitutes an invaluable tool
to pursue a range of projects for both galactic archaeology and exoplanet host star analyses. 
This sample is the second phase of the AMBRE project and will be followed by the parameterisation of the UVES archive
(Worley et al., 2014, in preparation).

\begin{sidewaystable*}
  \caption{Description of the columns in the table of HARPS stellar parameters delivered to ESO.}
  \label{tab:param-desc}
  \centering
  \begin{tabular}{lp{6.5cm}ccp{7cm}}
    \hline \hline
    Keyword & Definition & Value range & Null value & Determination \\ 
    \hline
    DP\_ID                        & ESO dataset identifier & --- & --- & FITS file header. \\
    OBJECT                        & Object designation as read in ORIGFILE & --- & --- & FITS file header. \\
    TARG\_NAME                    & Target designation as read in ORIGFILE & --- & --- & FITS file header. \\
    RAJ2000                       & Telescope pointing (right ascension, J2000) & --- & --- & Units = deg. FITS file header. \\
    DEJ2000                       & Telescope pointing (declination, J2000) & --- & --- & Units = deg. FITS file header. \\
    MJD\_OBS                      & Start of observation date & --- & --- & Units = Julian day. FITS file header. \\
    EXPTIME                       & Total integration time & --- & --- & Units = sec. FITS file header. \\
    S/N                           & Signal-to-noise ratio as estimated by the pipeline & $(10,\,\infty)$ & $\verb|NaN|$ & Internal routine. \\

    VRAD                          & Stellar radial velocity & $[-500,\, 500]$ & $\verb|NaN|$ & Units = km~s$^{-1}$.\\
    ERR\_VRAD                     & Error on the radial velocity & $(0,\,\infty)$ & $\verb|NaN|$ & Units = km~s$^{-1}$. FITS file header/AMBRE radial velocity calculation routine. \\
    VRAD\_CCF\_FWHM               & FWHM of the CCF between the spectrum and the binary mask & $(0,\, \infty)$ & $\verb|NaN|$ & FITS file header/AMBRE radial velocity calculation routine. \\
    VRAD\_FLAG                    & Quality flag on the radial velocity analysis & 0, 1, 2, 3, 4, 5 & \verb|NaN| & 0 = Excellet determination $\dots$ 5 = Poor determination, when $V_\mathrm{rad}$ determined by AMBRE pipeline. \verb|NaN| when $V_\mathrm{rad}$ determined by HARPS pipeline.\\

    TEFF                          & Stellar effective temperature ($T_\mathrm{eff}$) as estimated by the pipeline & $[4000,\, 7625]$ & $\verb|NaN|$ & Units = K. Null value used if $T_{\mathrm{eff}}$ outside accepted parameter limits.\\
    ERR\_INT\_TEFF                & Effective temperature internal error & $[10,\,100]$ & $\verb|NaN|$ & Units = K. Defined as function of S/N, it is equal to the 0.7 quantile to 
    the bin to which the S/N value belongs to; see Fig.~\ref{fig:repeat-error} and Sec.~\ref{sec:internal-err}.\\
    ERR\_EXT\_TEFF                & Effective temperature external error & $93$ & $\verb|NaN|$ & Units = K. Defined using the Porto sample; see Sec.~\ref{sec:extern-error-quant}. \\
 
    LOG\_G                        & Stellar surface gravity ($\log g$) as estimated by the pipeline & $[0,\,5.0]$ & $\verb|NaN|$ & Units = dex. Null value used if $\log g$ outside accepted parameter limits. \\
    ERR\_INT\_LOG\_G              & Surface gravity internal error & $[0.02, \,0.184]$ & $\verb|NaN|$ & Units = dex. See ERR\_INT\_TEFF for definition. \\
    ERR\_EXT\_LOG\_G              & Surface gravity external error & $0.26$ & $\verb|NaN|$ & Units = dex. See ERR\_EXT\_TEFF for definition.\\

    M\_H                          & Mean metallicity [M/H] as estimated by the pipeline & $[-3.5,\,1]$ & $\verb|NaN|$ & Units = dex. Null value used if [M/H] outside accepted parameter limits. \\
    ERR\_INT\_M\_H                & Mean metallicity internal error & $[0.01,\,0.07]$& $\verb|NaN|$ & Units = dex. See ERR\_INT\_TEFF for definition. \\
    ERR\_EXT\_M\_H                & Mean metallicity external error & $0.08$ & $\verb|NaN|$ &  Units = dex. See ERR\_EXT\_TEFF for definition.\\

    ALPHA                         & $\alpha$-elements over iron enrichment ([$\alpha$/Fe]) as estimated by the pipeline & $[-0.4,\,0.8]$ & $\verb|NaN|$ & Units = dex. Null value used if [$\alpha$/Fe] outside accepted parameter limits. \\
    ERR\_INT\_ALPHA               & $\alpha$-elements over iron enrichment internal error & $[0.005,\,0.048]$ & $\verb|NaN|$ & Units = dex. See ERR\_INT\_TEFF for definition.\\
    ERR\_EXT\_ALPHA               & $\alpha$-elements over iron enrichment external error & $0.04$ & $\verb|NaN|$ & Units = dex. See ERR\_EXT\_TEFF for definition. \\

    CHI2                          & $\log (\chi^2)$ of the fit between the observed and reconstructed synthetic spectrum at the \matisse\, parameters & $[-5,\,\infty)$ & $\verb|NaN|$ & Goodness of fit between final normalised and final reconstructed spectra.\\
    CHI2\_FLAG                    & Quality flag on the fit between the observed and reconstructed synthetic spectrum at the \matisse\, parameters & 0, 1, 2 & \verb|NaN| & 0 = Very good fit, 1 = Good fit, 2 = Poor fit \\

    ORIGFILE                      & ESO file name of the original spectrum being analysed & --- & --- & \\
    \hline
  \end{tabular}
\end{sidewaystable*}

\begin{table*}[h]
  \caption{Definition of the quality flags of the fit between the observed and the reconstructed synthetic
    spectra (see Table~3). In the second column, the flags are defined as: 0={\it Very Good}, 1={\it Good}, and 2={\it Acceptable}. 
    The third column refers to the percentage of spectra associated with the corresponding $\chi^2$ flag.
    The last column defines the selection criteria for the flags. $\chi^2_\mathrm{cold,fit}$ and $\sigma_\mathrm{cold}$; $\chi^2_\mathrm{warm,fit}$ 
    and $\sigma_\mathrm{warm}$; and $\chi^2_\mathrm{hot,fit}$ and $\sigma_\mathrm{hot}$ refer to 11 \%, 83 \%, and 2 \% of the delivered 
spectra, respectively.}
  \label{tab:chi2-flag}
  \centering
  \begin{tabular}{cccl}
    \hline \hline
    $T_\mathrm{eff}$ domain [K]& Flags & \% spectra & \multicolumn{1}{c}{$\chi^2$ limits} \\ 
    \hline
    \multirow{3}{*}{$4000 \leq T_\mathrm{eff} \leq 5000$} & 0 & 10 & $\chi^2 \leq \chi^2_\mathrm{cold,fit} (1 - \sigma_\mathrm{cold})$ \\
    & 1 & 59 & $\chi^2_\mathrm{cold,fit} (1 - \sigma_\mathrm{cold}) < \chi^2 \leq \chi^2_\mathrm{cold,fit}$  \\
    & 2 & 31 & $\chi^2_\mathrm{cold,fit} < \chi^2 \leq \chi^2_\mathrm{cold,fit} (1 + 0.5 \sigma_\mathrm{cold})$ \\
    \hline
    \multirow{3}{*}{$5000 < T_\mathrm{eff} \leq 6500$} & 0 & 44 & $\chi^2 \leq \chi^2_\mathrm{warm,fit}$ \\
    & 1 & 51 & $\chi^2_\mathrm{warm,fit} < \chi^2 \leq \chi^2_\mathrm{warm,fit}(1 + 1.5\sigma_\mathrm{warm})$  \\
    & 2 &  4 & $\chi^2_\mathrm{warm,fit}(1 + 1.5\sigma_\mathrm{warm}) < \chi^2 \leq \chi^2_\mathrm{warm,fit} (1 + 3\sigma_\mathrm{warm})$ \\
    \hline
    \multirow{3}{*}{$T_\mathrm{eff} > 6500$} & 0 & 26 & $\chi^2 \leq \chi^2_\mathrm{hot,fit} (1 - \sigma_\mathrm{hot})$ \\
    & 1 & 14 & $\chi^2_\mathrm{hot,fit} (1 - \sigma_\mathrm{hot}) < \chi^2 \leq \chi^2_\mathrm{hot,fit} (1 - 0.5\sigma_\mathrm{hot})$ \\ 
    & 2 & 60 & $\chi^2_\mathrm{hot,fit} (1 - 0.5\sigma_\mathrm{hot}) < \chi^2 \leq \chi^2_\mathrm{hot,fit}$\\ 
    \hline
  \end{tabular}
  \tablefoot{The percentages of the different temperature domains does not sum to 100 \% because we 
  have excluded spectra with $T_\mathrm{eff} < 4000$ K.}
\end{table*}

\begin{acknowledgements}
The AMBRE Project has been supported and would like to thank ESO, OCA and CNES. Most of the calculations have been performed with the high-performance computing facility SIGAMM, hosted by OCA. We sincerely acknowledge L. Pasquini for
initiating this project and M. Romaniello and J. Melnick
for their help within ESO. M.D.P. would like to thank the AMBRE 
team for the great period (A l'OCA Marco: Bouffe, Rugby, \'Etoiles). This research has made use of the SIMBAD database, operated at CDS, Strasbourg, France,  
of Astropy, a community-developed core Python package for Astronomy \citep{astropy2013} and matplotlib \citep{Hunter:2007}.
We thank the anonymous referee for her/his constructive comments that
improved the content of this article. 
\end{acknowledgements}

\bibliographystyle{aa}
\bibliography{depascale}
\end{document}